\documentclass[twocolumn,secnumarabic,amssymb, nobibnotes, aps, prd]{revtex4-1}

\usepackage[english]{babel}
\usepackage{bm}

\usepackage{amsbsy,mathtools} 

\newcommand{\ket}[1]{\left| #1 \right\rangle} 
\newcommand{\braket}[2]{\left\langle #1 \vphantom{#2} \right| \left. #2 \vphantom{#1} \right\rangle} 
\newcommand{\matrixel}[3]{\left\langle #1 \vphantom{#2#3} \right| #2 \left| #3 \vphantom{#1#2} \right\rangle} 

\newcommand{\chitwo}{$\chi^{(2)}$\,}
\newcommand{\hybrid}[1]{ h_{\xi}^{#1}}

\renewcommand{\v}[1]{\ensuremath{\bm{#1}}} 
\newcommand{\versor}[1]{\ensuremath{\bm{\hat{#1}}}} 

\newcommand{\vv}[1]{\ensuremath{\bar{\bm{#1}}}} 
\newcommand{\strain}{\vv{\varepsilon} }

\begin{document}

\title{Bond orbital description of the strain induced \\ second order optical susceptibility in silicon}%

\author{Pedro Damas}\email{pedro.damas@u-psud.fr}
\author{Delphine Marris-Morini}
\author{Eric Cassan}
\author{Laurent Vivien}\email{laurent.vivien@u-psud.fr}%

\affiliation{Institut d'Electronique Fondamentale, Universit\'e Paris Sud, CNRS, UMR 8622, Universit\'e Paris-Saclay, B\^at. 220, 91405 Orsay Cedex, France}
\date{09/11/2015}%
\begin{abstract}
We develop a theoretical model, relying on the well established sp3 \textit{bond-orbital theory}, to describe the strain-induced \chitwo in tetrahedrally coordinated centrosymmetric covalent crystals, like silicon. With this approach we are able to describe every component of the \chitwo tensor in terms of a linear combination of strain gradients and only two parameters $\alpha$ and $\beta$ which can be estimated theoretically. The resulting formula can be applied to the simulation of the strain distribution of a practical strained silicon device, providing an extraordinary tool for optimization of its optical nonlinear effects. By doing that, we were able not only to confirm the main valid claims known about \chitwo in strained silicon, but also estimate the order of magnitude of the \chitwo generated in that device.
\end{abstract}
\maketitle


\section{Introduction}

Silicon-based photonics has generated a strong interest in recent years, mainly for optical communications and optical interconnects in CMOS circuits. The main motivations for silicon photonics are the reduction of photonic system costs and the increase of the number of functionalities on the same integrated chip by combining photonics and electronics, along with a strong reduction of power consumption \cite{Fedeli2008}. However, one of the biggest constraints of silicon as an active photonics material is its vanishing second order optical susceptibility, the so called \chitwo, due to the centrosymmety of the silicon crystal. Without any second order nonlinear phenomena, fast and low power consumption optical modulation based on Pockels effect and wavelength conversions based on Second Harmonic Generation (SHG) are not possible in bulk Si \cite{Leuthold2010}. This is a very limiting factor when we expect silicon to be part of a solution to high performances and high energy efficient devices \cite{Reed2010}.

To overcome this problem, strain has been used as a way to deform the crystal and destroy the centrosymmetry which inhibits \chitwo . In fact, over the last few years Pockels electro-optic modulation \cite{Jacobsen2006,Chmielak2011a,Chmielak2013,Puckett2014,Damas2014} and SHG \cite{Cazzanelli2012,Schriever2012} have been claimed to be demonstrated in devices where the silicon active region is strained by a stress overlayer, usually made of SiN. Motivated by its enormous potential, the interest in strained silicon photonics devices has been growing in the past years. However, there is a lack of fundamental understanding on the process through which the strain tensor \strain generates non vanishing \chitwo tensor components. In other words, there is no available general quantitative relationship between these two quantities.

Despite the different attempts to find a solution to this problem, no theoretical model showing a practical relationship between the components of the second order nonlinear optical susceptibility \chitwo and the strain tensor \strain has been published yet. Such model is fundamental in the field of strained silicon photonics because it permits a connection between the strain effects, easy to simulate with the right computational tools, and the respective induced nonlinear phenomena. Without this knowledge, it is not possible to design and optimize structures based on strained silicon structures for the maximum outcome of \chitwo effects.

The first proposed models connecting \chitwo with \strain \cite{Govorkov1989, Huang1994} were based on the \textit{deformation potentials in semiconductors}. The deformation potentials theory relies on the de-localisation of the Bloch wavefunction over the entire crystal and proves to be a good description for the study of transport properties of electrons. However, it is known now that the nonlinear optical properties in covalent crystals are mainly due to the properties of the localized electrons in the covalent bonds between the different atoms of the crystal \cite{Levine1973,Levine1969,Kleinman1962a,Harrison1974,Aspnes2010}. Thus, theories based on the deformation potentials not only prove to be very limiting for extracting the different \chitwo components in terms of \strain , but also do not show a very good numerical agreement with the experimental data for \chitwo in strained silicon \cite{Govorkov1989,Hon2009a}.

Later, simpler models based on Coulomb interactions between atoms were also suggested \cite{Hon2009, Hon2009a}, but none of them proved to a good description of the phenomena, both numerically and conceptually. To overcome this difficulty, \textit{ab-initio} calculations were performed as an attempt to understand how the change in position of the atoms enables \chitwo in the crystal \cite{Cazzanelli2012,Luppi2015}. Although they proved to be an accurate description of the problem, they are very computationally demanding and do not provide a practical, quantitative and spatial relationship between \chitwo and the strain \strain in the crystal and thus do not allow for device design over the strain distribution.

Even though the underlying process has not been described yet, it has been widely claimed that there is a direct relationship between \chitwo and the strain gradients inside silicon \cite{Cazzanelli2012,Chmielak2013,Puckett2014,Damas2014,Schriever2015}.  In fact, very recently Manganelli et al. \cite{Manganelli2015} proposed a model based on symmetry arguments to show a linear relationship between both the spatial distribution of \chitwo and \strain tensor components. However, even though the theory is well developed, it is very dependent on parameters required to be determined experimentally. This poses a problem because very recently it has been shown that the reported electro-optic measurements of strained silicon devices have a strong contribution from free carriers effects inside the silicon waveguide \cite{Azadeh2015,Sharma2015,Schriever2015}. Therefore, most of the available numerical data of strain induced \chitwo in silicon waveguides available in the literature was misinterpreted and discredited, making it impossible at the moment to find the experimental parameters predicted by the model presented in \cite{Manganelli2015}.

Although many works have been reported, no consistent study based on the \textit{bond orbital model} has been reported yet on the study of the strain induced nonlinear effects in Si. The \textit{bond orbital model} is the most widely accepted model to describe the properties of electrons in the bonds of a tetraheral covalent crystals like silicon \cite{HarrisonBook1989, Harrison1973,Harrison1974}. Relying on the well established quantum mechanics of the sp3 hybridization of silicon valence electrons, the bonding electrons are treated quantum-mechanically to describe their properties. This procedure has already been used to characterize the strain effects \cite{Harrison1974b} and nonlinear optical properties \cite{HarrisonBook1989} in covalent crystals, although treated separately.

In the present work, we rely on the original work first developed by Harrison et al. in \cite{Harrison1973} to describe the bonding electrons. By using the \textit{bond orbital model} and keeping only first order effects in strain and strain gradients, we are able to show a practical way of knowing any \chitwo component generated by the strain (\strain) in that point in space. The results not only show very good agreement with the main claims made in the literature about the relationship between \chitwo and strain in silicon, but also depends only on 2 parameters which can be predicted theoretically and depend only on the material.

The organization of this paper is as follows: we start by presenting the general reasoning behind the model before entering into the quantum mechanical treatment of the sp3 hybridization in strained covalent crystals, in part \ref{sec:quantum mechanics}. In section \ref{sub:polarity} we proceed to the calculation of the strain induced polarity of a covalent bond and in section \ref{sub:dipole}, we deduce its second order nonlinear dipole moment. This will allow us to extract the strain induced \chitwo components in terms of the strain tensor (\strain) components. Finally, in section \ref{sec:evaluation} we apply our model to a particular example device, showing the degree of the agreement and its practical applicability.

\section{The strain induced \chitwo \,in covalent crystals}

\begin{figure}
\includegraphics[width=0.45\textwidth]{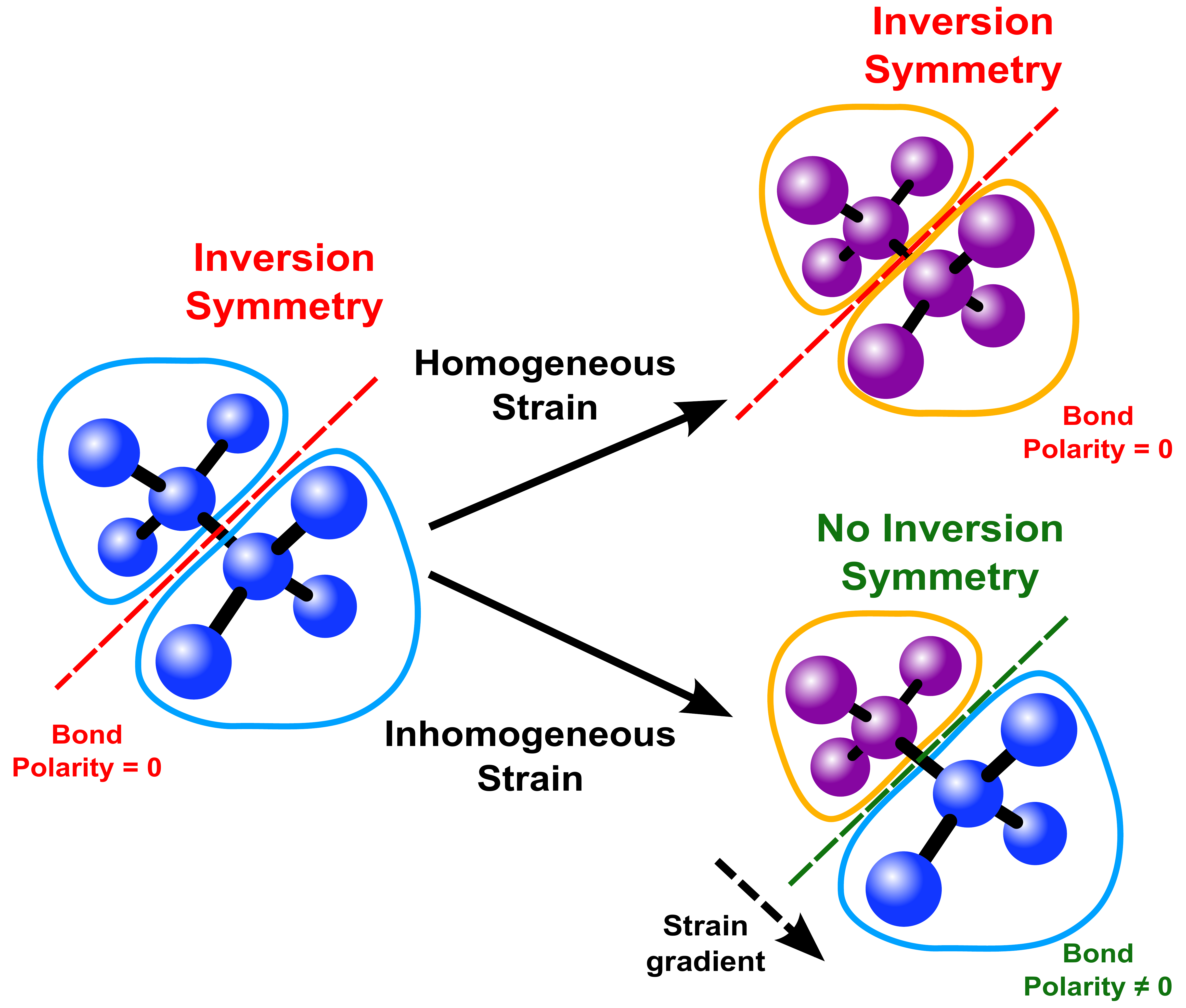} 
\caption{ The representation of the process of generation bond polarity, where different contour colours represent different electron energies. An homogeneous strain changes the energy of the bonding electrons, but it is still the same in both sides of the bond (yellow contours), keeping its inversion symmetry. However, a strain gradient, changes the energy in both sides of the bond (represented with the yellow and blue contours), creating a difference in energy polarity and destroying the inversion symmetry. The generated bond polarity is the origin of \chitwo in that bond. \label{fig:strainProcess}}

\end{figure}

Silicon is a tetrahedral covalent crystal, where two neighbour atoms are bond together by sharing electrons, creating a \textit{covalent bond} \cite{HarrisonBook1989}. Four covalent bonds organize themselves in a tetrahedral configuration as shown in Fig.\ref{fig:strainProcess}. Understanding any property of the crystal (like \chitwo ) requires studying the quantum mechanical interactions of the electrons in the bonds. However, to make it clearer for the reader, before entering in any mathematical description, we start by presenting the process suggested in this work for the generation of \chitwo in each bond due to the strain.

It is known that \chitwo depends primarily on the polarity of the bonds of a crystal \cite{Harrison1974,Levine1973,Levine1969}. The polarity of a bond is the difference of energy of the two electrons making the bond. In a centrosymmetric crystal, since the electrons feel the same energy in both directions, the bonds are unpolar, as shown in Fig. \ref{fig:strainProcess}. However, when strain is applied to a crystal, the atomic configuration changes and a bond becomes polar if and only if there is a strain gradient in the direction of the bond. This process is presented schematically in Fig. \ref{fig:strainProcess}, where the blue and yellow contours represent different values of energy. This explains why inhomogeneous strain fields are required to induce bond polarity and thus \chitwo, because homogeneous strain changes the electronic energy in the same way in both electrons of the bond.

This is the idea behind the mathematical and quantum description of the problem: first we calculate how the strain changes the polarity of a bond and from then we deduce how that strain induces second order nonlinear effects. Moreover, even though in this work we focus on silicon atoms, this procedure can be applied to any covalent diamond crystal structure.

\subsection{Quantum mechanical treatment of a strained covalent crystal} \label{sec:quantum mechanics}

\begin{figure}
\includegraphics[width=0.45\textwidth]{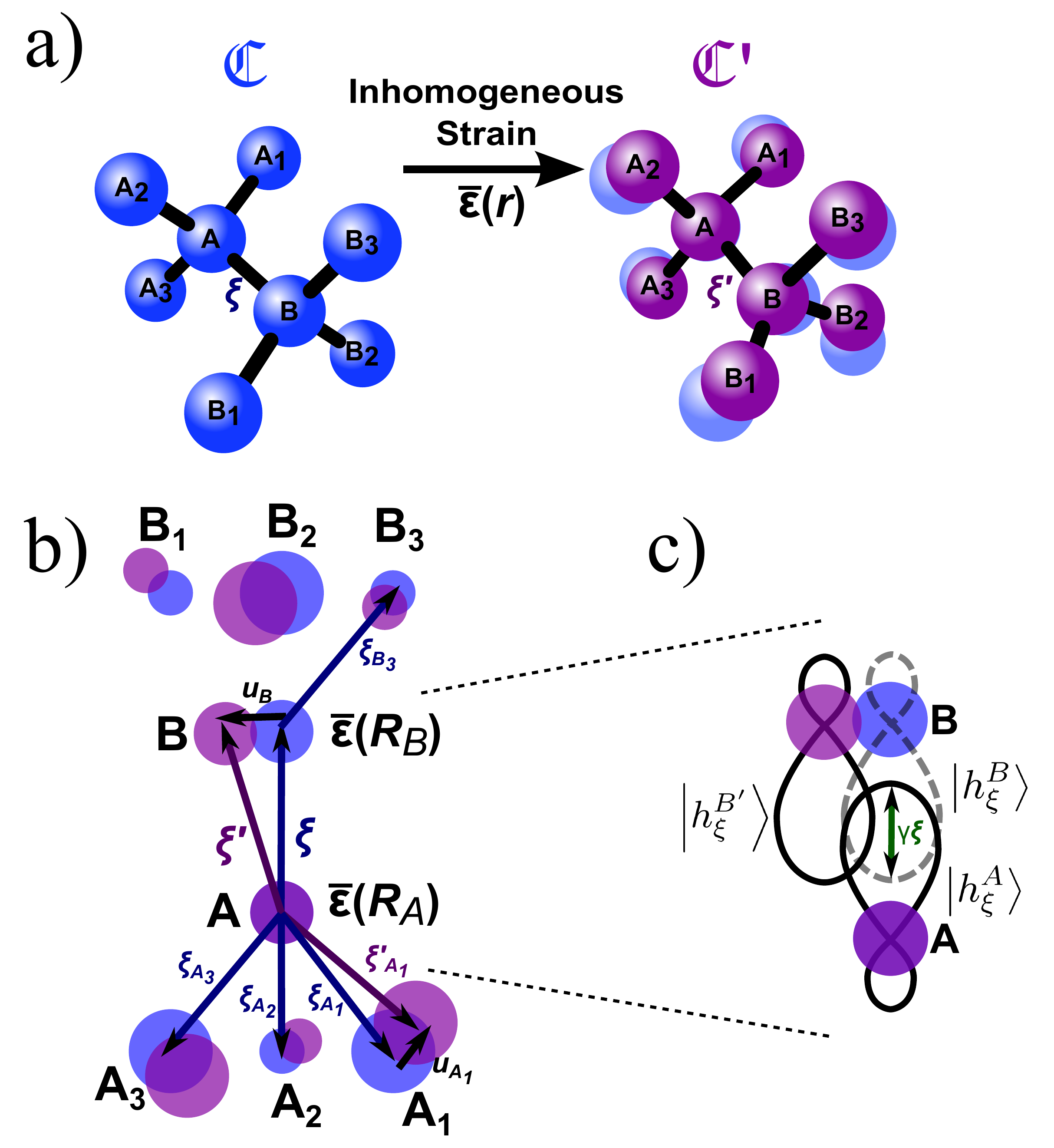} 
\caption{ a) 3D Representation of the bond $\xi$ between the general atoms $A$ and $B$ in a unstrained silicon lattice $\mathfrak{C}$ unit cell, in blue. In purple is the corresponding structure when an inhomogeneous strain field is applied, generating the lattice $\mathfrak{C'}$. b) The 2D projection of the unit cells in a), with the representation of the relevant vectors used in the text. c) Schematic representation of the bond hybrids in the original and in the strained crystal.
\label{fig:strainedHybrids}}
\end{figure}

Consider a tetrahedral covalent crystal $\mathfrak{C}$, represented in blue in the 3D scheme of Fig. \ref{fig:strainedHybrids} a). The four valence electrons in a general Si atom $A$ organize themselves in four different sp3 hybrids $\ket{h_{\xi}^A}$ pointing in the direction of the four nearest neighbours, creating four bonds $\xi \;(\xi = 1,2,3,4)$. Such hybrid orbitals on any given atom are orthogonal to each other if the near neighbours are exactly tetrahedral, like in an unstrained Si crystal \cite{Harrison1973}.

Consider now the lattice $\mathfrak{C}'$ (represented in purple), the strained version of $\mathfrak{C}$. Each atom $n$ in $\mathfrak{C}'$ is slightly moved by a vector $\v{u}_n$ in relation to $\mathfrak{C}$ and the new atomic organization in $\mathfrak{C}'$ will change a bond $\xi$ into $\xi'$ (see Fig. \ref{fig:strainedHybrids} b) ). To study this new bonding, it may sound appealing to construct four new hybrids out of the atomic orbitals, with the direction $\xi'$. However, since the atoms are not arranged in a tetrahedral configuration anymore, such set of hybrids would not be orthogonal and it would require a special treatment afterwards \cite{Harrison1973a}. To overcome this problem, we will construct the wavefunction of the electrons in any atom $n$ has a combination of the original hybrids $\ket{h_{\xi}^n}$, placed in the new atomic positions. This is schematically represented in the Fig. \ref{fig:strainedHybrids} c), making it possible to deal with the quantum mechanical subtleties of this problem, as it will be apparent later on.

%
%
Consider now, without any loss of generality, atom $A$, taken as the atom which preserves the same position in $\mathfrak{C}'$ and $\mathfrak{C}$, as shown in Fig. \ref{fig:strainedHybrids} a) and b). This atom is connected to four other atoms. We focus on one of its four bonds, connecting atom $A$ with atom $B$ and call it bond $\xi$. The bond vector $\v{\xi}$ (associated with bond $\xi$) is defined as the vector from atom $A$ to atom $B$. (Fig. \ref{fig:strainedHybrids} b) ).

To study the quantum mechanical properties of the electrons in the bond, we must start by building its Hamiltonian. The one-electron Hamiltonian of the strained crystal lattice $\mathfrak{C}'$ is given by \cite{Huang1975}:
\begin{equation}\label{eq:H'}
H' = T + \sum_n V'_n = T + V'_A + V'_B + \sum_{n\neq A,B} V'_n 
\end{equation}
where $T$ is the kinetic energy of the electron and $V'_n(\v{r}) = V(\v{r} - \v{R}'_n)$ is the potential due to the atom in position $\v{R}_n'$. This potential can be written as
\begin{equation}\label{eq:DeltaV}
V'_n(\v{r}) = V_n(\v{r})+\Delta V_n(\v{r})\; ,
\end{equation}
with $\Delta V_n$ being the contribution from the strain effects, vanishing for an unstrained crystal. We have explicitly separated $V'_A$ and $V'_B$ from the sum in eq. \ref{eq:H'} because we are focusing on the bond between atoms $A$ and $B$ and its treatment is more clear this way.

The matrix element of $H'$ in $\ket{\hybrid{A}}$ is given by
\begin{multline}\label{eq:H' explicit}
\matrixel{ \hybrid{A} }{H'}{ \hybrid{A} } = \matrixel{ \hybrid{A} }{T + V'_A}{ \hybrid{A} } + \\ + \matrixel{ \hybrid{A} }{V'_B}{ \hybrid{A} } + \sum_{n\neq A,B} \matrixel{ \hybrid{A} }{V'_n}{ \hybrid{A} }
\end{multline}
%
%
being equivalent in $\ket{\hybrid{B'}}$. We should now try to relate them to the same matrix elements of the unstrained Hamiltonian $H=T+\sum_n V_n$ in the original  hybrids basis. Since the hybrid wavefunctions $\ket{\hybrid{A}}$ and $\ket{\hybrid{B'}}$ are respectively centred at the atoms $A$ and $B$ in $\mathfrak{C}'$, we have:
\begin{eqnarray}
\matrixel{ \hybrid{A} }{T + V'_A}{ \hybrid{A} } &=& \matrixel{ \hybrid{A} }{T + V_A}{ \hybrid{A} } \\
\matrixel{ \hybrid{B'} }{T + V'_B}{ \hybrid{B'} } &=& \matrixel{ \hybrid{B} }{T + V_B}{ \hybrid{B} }
\end{eqnarray}

Moreover, because of the symmetry of the bond, the potential of $A$ in $\ket{\hybrid{B'}}$ is the same as the potential of $B$ in $\ket{\hybrid{A}}$. Thus,
\begin{equation}
\matrixel{ \hybrid{A} }{V'_B}{ \hybrid{A} } = \matrixel{ \hybrid{B'} }{V'_A}{ \hybrid{B'} } = \epsilon_{AB}+\Delta \epsilon_{AB}
\end{equation}
where $\epsilon_{AB}=\matrixel{ \hybrid{A} }{V_B}{ \hybrid{A} }$ and $\Delta \epsilon_{AB}$ is the correction accounting for the new relative position of atoms $A$ and $B$ in $\mathfrak{C'}$.

For all the other atoms ($n\neq A,B$), we can write:
\begin{eqnarray}
\matrixel{ \hybrid{A} }{ V'_n }{ \hybrid{A} }&=& \matrixel{ \hybrid{A} }{ V_n + \Delta V_n }{ \hybrid{A} }= \nonumber \\ 
&=&   \matrixel{ \hybrid{A} }{ V_n }{ \hybrid{A} } +  \matrixel{ \hybrid{A} }{\Delta V_n }{ \hybrid{A} } \\
\matrixel{ \hybrid{B'} }{ V'_n }{ \hybrid{B'} }&=& \matrixel{ \hybrid{B'} }{ V_n + \Delta V_n }{ \hybrid{B'} }= \nonumber \\ 
&=&   \matrixel{ \hybrid{B} }{ V_n }{ \hybrid{B} } +  \matrixel{ \hybrid{B'} }{\Delta V_n }{ \hybrid{B'} } 
\end{eqnarray}
From the previous analysis and after eq. \ref{eq:H' explicit}, we are in conditions of writing the matrix elements of $H'$ as
\begin{eqnarray*}
\matrixel{ \hybrid{A} }{H'}{ \hybrid{A} } &=& \epsilon_A + \Delta \epsilon_{AB} + \sum_{n\neq A,B} \matrixel{ \hybrid{A} }{\Delta V_n }{ \hybrid{A} } \\
\matrixel{ \hybrid{B'} }{H'}{ \hybrid{B'} } &=& \epsilon_B + \Delta \epsilon_{AB} + \sum_{n\neq A,B} \matrixel{ \hybrid{B'} }{\Delta V_n }{ \hybrid{B'} } \\
\matrixel{ \hybrid{A} }{H'}{ \hybrid{B'} } &=& U'_{AB}
\end{eqnarray*}
The terms $\matrixel{ \hybrid{A} }{H}{ \hybrid{A} } = \epsilon_A$ and $\matrixel{ \hybrid{B} }{H}{ \hybrid{B} } = \epsilon_B$ are the average energy of $\ket{\hybrid{A}}$ and $\ket{\hybrid{B}}$ in $\mathfrak{C}$ and because of its centrosymmetry, they have both the same value $\epsilon_A = \epsilon_B$.

In addition, the effects of strain $\varepsilon$ on the cross term $U'_{AB}$ has been studied by Harrison et al. in \cite{Harrison1974b} and it is shown that
\begin{equation*}
U'_{AB} =  U_{AB}(1 - \eta \varepsilon^2)
\end{equation*}
where $U_{AB} = \matrixel{ \hybrid{A} }{H}{ \hybrid{B} }$ and $\eta$ a fitting constant. This shows that the Hamiltonian cross term has a second order correction in the strain effects.

The obtained Hamiltonian matrix elements will be used now to calculate the polarity of the bond in the strained crystal.

\subsection{Strain induced bond polarity}	\label{sub:polarity}
The polar energy (or polarity) of the bond $\xi$ in the strained crystal is defined by \citep{Harrison1973,Harrison1974}
\begin{eqnarray}
\sigma &=& \frac{\matrixel{ \hybrid{A} }{H'}{ \hybrid{A} } - \matrixel{ \hybrid{B'} }{H'}{ \hybrid{B'} }}{2} = \\
&=& \sum_{n\neq A,B} \frac{ \matrixel{ \hybrid{A} }{ \Delta V_n}{ \hybrid{A} } - \matrixel{ \hybrid{B'} }{ \Delta V_n}{ \hybrid{B'} } }{2} \label{eq:polarity definition}
\end{eqnarray}
and it can be immediately seen that if $\Delta V_n = 0$, i.e. no strain is applied to the crystal, $\sigma = 0$ and the bonds are non-polar. This is the reason behind the vanishing \chitwo in non-strained centrosymmetric crystals.

By reducing the sum in equation \ref{eq:polarity definition} only to the interaction between the first neighbours of atoms $A$ and $B$ individually, $A_i$ and $B_i$ $i=1,2,3$ respectively as shown in Fig. \ref{fig:strainedHybrids} a) and b), equation \ref{eq:polarity definition} reduces to
\begin{equation}\label{eq:reduced polarity}
\sigma = \sum_{i=1}^3 \frac{ \matrixel{ \hybrid{A} }{ \Delta V_{A_i}}{ \hybrid{A} } - \matrixel{ \hybrid{B'} }{ \Delta V_{B_i}}{ \hybrid{B'} } }{2}
\end{equation}

Despite we have not said anything about the form of the crystal potential $V(\v{r})$ yet, we know it is a central potential ($V(\v{r}) = V(r)$) and for $r$ big enough, it should behave like a Coulomb potential. Therefore, for small displacements of the atoms, we may assume that $ \lVert \v{u}_n \rVert\ll \lVert\v{r} - \v{R}_n \rVert$ and the form of $\Delta V_n(\v{r})$ defined in equation \ref{eq:DeltaV} can be taken by performing a first order Taylor expansion of the potential $V'_n(\v{r})$: 
\begin{eqnarray*}
V'_n(\v{r}) &=& V(\v{r} - \v{R}'_n) = V(\v{r} - \v{R}_n-\v{u}_n) \sim \\
&\sim & V(\v{r} - \v{R}_n)- \nabla V(\v{r} - \v{R}_n)\cdot \v{u}_n
\end{eqnarray*}
By defining $\nabla V_n \equiv \nabla V(\v{r} - \v{R}_n)$, it is clear that
\begin{equation}\label{eq:DVn2}
\Delta V_{n} = -\nabla V_n\cdot \v{u}_{n} 
\end{equation}
which can only be evaluated once we know the explicit form of $V(\v{r})$.

Using this definition along with the symmetries of the bond, the central properties of the potential $V(\v{r})$ and bearing in mind that the hybrid wavefunctions statisfy $\hybrid{B'}(\v{r}) = \hybrid{A}(\v{R}'_B-\v{r})$, it can be shown that
\begin{eqnarray*}
\matrixel{ \hybrid{A} }{ \Delta V_{A_i}}{ \hybrid{A} } &=& -\matrixel{ \hybrid{A} }{ \nabla V_{A_i}}{ \hybrid{A} }\cdot \v{u}_{A_i} \\
\matrixel{ \hybrid{B'} }{ \Delta V_{B_i}}{ \hybrid{B'} } &=& \matrixel{ \hybrid{A} }{ \nabla V_{A_i}}{ \hybrid{A} }\cdot (\v{u}_{B_i} - \v{u}_{B})
\end{eqnarray*}
leading to the simplification of equation \ref{eq:reduced polarity} into
\begin{equation}\label{eq:polarity2}
\sigma = \sum_{i=1}^3 \frac{ \matrixel{ \hybrid{A} }{ \nabla V_{A_i}}{ \hybrid{A} } \cdot (\v{u}_{B} - \v{u}_{B_i} -\v{u}_{A_i})}{2}
\end{equation}

Assume now that there is a known strain field $\strain(\v{r})$ in the crystal as shown in Fig. \ref{fig:strainedHybrids} b) and that the strain applied to each atom $n$ is given by $\strain(\v{R}_n)$. From simulations of the strain distribution in a Si crystal, it can be shown that the strain is a slowly varying function over the bond length $d$ ($\frac{\partial \varepsilon}{\partial x} . d < 10^{-6}$). In that case, by defining $\v{\xi}_{A_i}$ as the bond vector between atoms $A$ and $A_i$ and respectively for B (Fig. \ref{fig:strainedHybrids} b) ), it is easy to show using elasticity theory that \cite{Landau7,BirSymmetry}
\begin{eqnarray}
\v{u}_{A_i} &\simeq& \strain(\v{R}_A)\cdot \v{\xi}_{A_i} \label{eq:uA}\\
\v{u}_{B_i} &\simeq& \v{u}_B + \strain(\v{R}_B)\cdot \v{\xi}_{B_i} \label{eq:uB}
\end{eqnarray}

Putting together equations  \ref{eq:polarity2}, \ref{eq:uA}, \ref{eq:uB} and bearing in mind that $\v{\xi}_{B_i} = - \v{\xi}_{A_i}$ (Fig. \ref{fig:strainedHybrids} b)), we deduce
\begin{equation*}
\sigma_{\xi} = \frac{1}{2}\sum_{i=1}^3 \matrixel{ \hybrid{A} }{ \nabla V_{A_i}}{ \hybrid{A} }\cdot \left[ \strain(\v{R}_B) - \strain(\v{R}_A)\right]\cdot \v{\xi}_{A_i}
\end{equation*}

Since the strain changes slowly in distances of the bond length $d$, we can relate the component $kl$ of the strain tensor in atoms $A$ and $B$ by making a first order Taylor expansion
$$\varepsilon_{kl}(\v{R}_B) = \varepsilon_{kl}(\v{R}_A+\v{\xi}) \sim \varepsilon_{kl}(\v{R}_A)+\nabla\varepsilon_{kl}(\v{R}_A) \cdot \v{\xi}$$
which allows us to write the final expression of the polarity of a bond $\xi$ in atom $A$ as
\begin{equation} \label{eq:polarity final1}
\sigma_\xi(\v{R}_A) = \frac{1}{2} \sum_{i=1}^3 \v{\theta}^{\xi}_i \cdot \vv{\Xi}(\xi;\v{R}_A) \cdot \v{\xi}_{A_i}
\end{equation}
where we have defined the rank-2 tensor $\vv{\Xi}(\xi;\v{R})$ (related to the bond $\xi$ in the atom located in the position $\v{R}$), whose component $kl$ is given by
\begin{equation}\label{eq:XI}
\Xi(\xi;\v{R})_{kl} = \left.\left( \frac{\partial \varepsilon_{kl}}{\partial x},
\frac{\partial \varepsilon_{kl}}{\partial y},
\frac{\partial \varepsilon_{kl}}{\partial z} \right)\right|_\mathbf{R} \cdot \v{\xi}
\end{equation}

\begin{figure}
\includegraphics[width=0.45\textwidth]{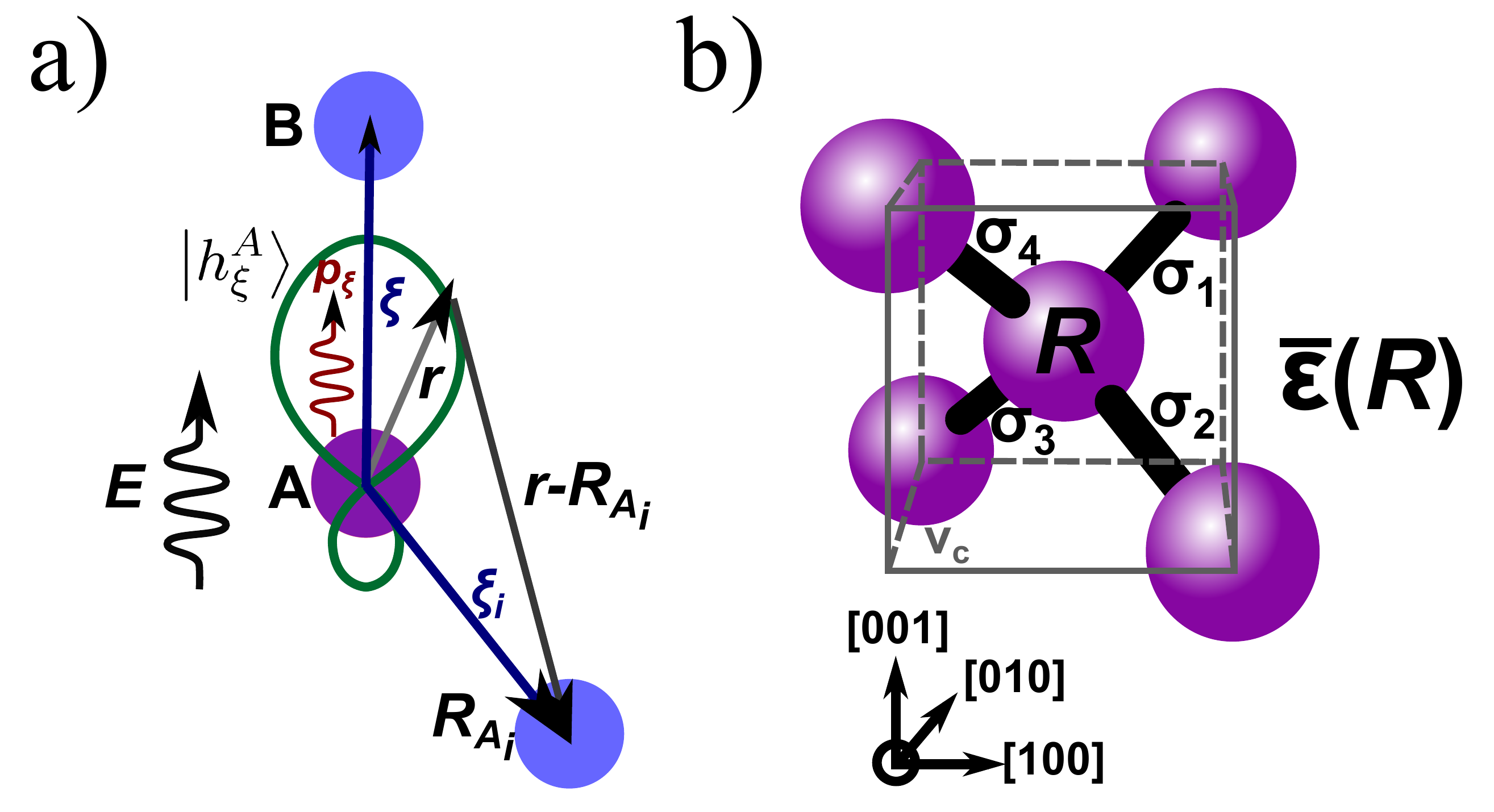} 
\caption{a) Representation of the optical electrid field $\v{E}$ generating the bond dipole moment $\v{p_{\xi}}$ and the vectors representation of the atomic position $\v{R}_{A_i}$ of atom $A_i$ and electron position $\v{r}$, useful for the integral in eq. \ref{eq:theta2}. 
b)Silicon unit cell (with volume $v_c$) centred at the position $\v{R}$ where each bond has its own polarity $\sigma_i$ induced by the strain tensor $\strain(\v{R})$ in the center of that unit cell.
\label{fig:UnitCellPol}}

\end{figure}

Furthermore, the vector $\v{\theta}^{\xi}_i$ is defined by
\begin{eqnarray}\label{eq:theta def}
\v{\theta}^{\xi}_i &=& \matrixel{ \hybrid{A} }{ \nabla V_{A_i}}{ \hybrid{A} } = \\ 
&=& \int_\infty |h_{\xi}^A(\v{r})|^2 \cdot \nabla V(\v{r}-\v{R}_{A_i})dV = \\
&=& \int_\infty |h_{\xi}^A(\v{r})|^2 \cdot \left.\frac{\partial V}{\partial r}\right\vert_{\v{r}-\v{R}_{A_i}} \cdot \frac{ \v{r}-\v{R}_{A_i} }{\lVert \v{r}-\v{R}_{A_i} \rVert} dV \label{eq:theta2}
\end{eqnarray}
Its evaluation can be done with the help of the scheme in in Fig. \ref{fig:UnitCellPol} a). It does not depend on the atom $A$ in particular, but only on the unstrained bonds $\v{\xi}_i \equiv \v{\xi}_{A_i}, \; i=1,2,3$, which are the bonds different than $\xi$ in the unit cell (compare Fig. \ref{fig:UnitCellPol} a) with Fig. \ref{fig:strainedHybrids} b)).
The closed form of $\v{\theta}^{\xi}_i$ can only be found once the potential $V(\v{r})$ is known. However, regardless of that, we can always define
\begin{equation}\label{eq:alphaBeta}
\v{\theta}^{\xi}_i = \alpha \v{\xi} + \beta \v{\xi}_i
\end{equation}
where $\alpha$ and $\beta$ are parameters whose values are related to the projections of $\v{\theta}^{\xi}_i$ on $\v{\xi}$ and $\v{\xi}_i$ respectively and are characteristic of the crystal material in consideration. This definition together with equation \ref{eq:polarity final1}, allows us to write 
\begin{equation} \label{eq:polarity final}
\sigma_\xi(\v{R}) = \frac{1}{2} \sum_{i=1}^3 \left[\alpha \v{\xi} + \beta \v{\xi}_i\right] \cdot \vv{\Xi}(\xi;\v{R}) \cdot \v{\xi}_i \quad .
\end{equation}

Expression \ref{eq:polarity final} is the final expression for the polar energy (or polarity) of any bond $\xi$ in an atom centred in the unit cell located in $\v{R}$ (see Fig. \ref{fig:UnitCellPol} b) ). The subscript $\xi$ identifies one of the four bonds, which defines the corresponding bond vector $\v{\xi}$ and then the 3 other vectors $\v{\xi}_i,\, i=1,2,3$. We see that it depends on the strain gradients through $\vv{\Xi}(\xi;\v{R})$, which is non-zero only if there is a strain gradient component in the direction of the bond $\xi$. This is relevant because it shows that in a centro-symmetric crystal not only a strain gradient is required to create a polar bond, but it also gives the preferred gradient direction to obtain maximum polarity in bond $\v{\xi}$.

Moreover, once the strain distribution $\vv{\varepsilon}(\v{r})$ is known, the only parameters left to know are $\alpha$ and $\beta$. These two coefficients are the only unknowns of the model presented so far and their value (defined in equation \ref{eq:alphaBeta}) should be found experimentally, but this particular point requires further attention and it will be discussed later in section \ref{sec:numerical}.
\subsection{The second order nonlinear optical dipole moment} \label{sub:dipole}
Now that the polarity of a strained bond is known in terms of the strain tensor \strain, we can explore the nonlinear optical properties of the bond by computing the bond wavefunction and extract the second order dipole moment of the electrons in that bond. We now approach the problem as in the original theory done by Harrison et al in \cite{Harrison1974}. The bond wavefunction $\ket{b_{\xi}}$ of the bond ${\xi}$ is considered to be a combination of the two adjacent hybrids of the atoms forming that bond (Fig. \ref{fig:strainedHybrids} c) ) \cite{Harrison1973a, Harrison1974, Harrison1973}
\begin{equation}
\ket{b_{\xi}} = u_A \ket{\hybrid{A}} + u_B \ket{\hybrid{B'}}
\end{equation}
and is obtained by minimizing the bond energy $\epsilon_B = \matrixel{b_{\xi}}{H'}{b_{\xi}}/\braket{b_{\xi}}{b_{\xi}}$. In doing so, we are implicitly neglecting all the matrix elements of the Hamiltonian $H'$ and all other hybrid overlaps are neglected (or absorbed in the parameters we have retained). It is important to bear in mind that $\braket{\hybrid{A}}{\hybrid{B'}}=S\neq0$. Under these conditions, the explicit values for $u_A$ and $u_B$ can be found in the original work presented in \cite{Harrison1974}.

The average position of the bond wavefunction $\ket{b_{\xi}}$ in a strained crystal, in relation to the center of the bond, can be shown to be given by \cite{Harrison1974}
\begin{equation}
\langle \v{r} \rangle = \matrixel{b_{\xi}}{\v{r}}{b_{\xi}} = \left( u_B^2 - u_A^2 \right) \left[  \left(\gamma \vv{1} + \vv{\varepsilon} \right)\cdot \frac{\v{\xi}}{2} \right]
\end{equation}
which reduces to $\left( u_B^2 - u_A^2 \right) \gamma \v{\xi}/2$ in a non-strained crystal, the same result presented in \cite{Harrison1974}. In its original work, Harrison \cite{Harrison1974} introduces the parameter $\gamma$ which accounts for the distance between the "center of gravity" of each hybrid as shown in Fig. \ref{fig:strainedHybrids} c) and it would be unity if they were centred at the nucleus.

When an optical electric field $\v{E}$ interacts with the bond, it will induce a dipole moment $\v{p}_{\xi}$, as represented in red in Fig. \ref{fig:UnitCellPol} a). That dipole moment will change the Hamiltonian by a term $\Delta H = -\v{p} \cdot\v{E} =-2e\v{r}\cdot\v{E} $, which will change the bond wavefunction $\ket{b_{\xi}}$, yielding new coefficients $u_A$ and $u_B$. Finally, the dipole moment in the bond, created by the optical field will be given by $\v{p}  =\langle \v{p} \rangle_{\xi} = -2e\langle \v{r} \rangle_{\xi} =  -2e\matrixel{b_{\xi}}{\v{r}}{b_{\xi}}$ which will depend on the intensity of the optical electric field $\v{E}$. By expanding $\langle \v{p}_{\xi} \rangle$ in powers of $\v{E}$, the second order term is shown to be given by \cite{Harrison1974,HarrisonBook1989}
\begin{multline}
\v{p^{(2)}_{\xi}} = -3\left(\frac{e}{2}\right)^3 \left[ \frac{\sigma_{\xi}U^{'2}_{AB}(1-S^2)}{\left( \sigma_{\xi}^2(1-S^2)+U^{'2}_{AB}\right)^{\frac{5}{2}}} \right] \cdot \\
\cdot \left[ \left(\gamma \vv{1} + \vv{\varepsilon} \right)\cdot \v{\xi} \cdot \v{E} \right]^2 \cdot \left[  \left(\gamma \vv{1} + \vv{\varepsilon} \right)\cdot \v{\xi} \right] \label{eq:dipole1}
\end{multline}
The previous expression is drastically simplified if we stick only to first order terms in strain effects, i.e. in $\partial\varepsilon / \partial x$ and $\varepsilon$. Since $\sigma_{\xi}$ defined in \ref{eq:polarity final} is directly proportional to the strain gradients, keeping the terms proportional to \strain in eq. \ref{eq:dipole1} means retaining terms of the form $\varepsilon . \frac{\partial \varepsilon}{\partial x}$, i.e. of second order in strain effects. Therefore, we must remove all \strain terms from eq. \ref{eq:dipole1} to keep everything to first order in strain effects. Moreover, $U_{AB}$ is nothing more than half of the bandgap $E_g/2$ of the crystal \cite{Harrison1974,Jha1968,HarrisonBook1989,Huang1994,Philips1968a} which is much bigger than the strained induced polarity of the bond, $\sigma_{\xi}$. All these considerations finally lead to:
\begin{equation}\label{eq:final dipole}
\v{p^{(2)}_{\xi}} \simeq -3 \left( \frac{e\gamma}{E_g} \right)^3 \sigma_{\xi} (1-S^2) \left( \v{\xi} \cdot \v{E} \right)^2 \cdot \v{\xi}
\end{equation}

Expression \ref{eq:final dipole} is the final second order nonlinear dipole moment to first order in the strain effects. It can be seen the direct relationship between $\sigma_{\xi}$ and $\v{p^{(2)}}$, which justifies why nonpolar bonds do not contribute to 2\textsuperscript{nd} order nonlinear effects. Also, it is worth to stress the fact that the strain effects in $\v{p^{(2)}_{\xi}}$ are all inside $\sigma_{\xi}$: everything else, including the bond vector $\v{\xi}$, is related to the unstrained crystal lattice. This is relevant because it shows that to first order in the strain effects, only the strain gradients are important to the polarity of the bonds and thus to \chitwo generation.

The macroscopic 2\textsuperscript{nd} order nonlinear polarization $\v{P}^{(2)}(\v{r})$ is the sum of the contributions of the 4 bonds in the crystal unit cell centred in $\v{r}$, divided by its volume (Fig. \ref{fig:UnitCellPol} b) ). Thus
\begin{eqnarray}
\v{P^{(2)}} &=& \frac{1}{v_c} \sum_{\xi=1}^4 \v{p^{(2)}_{\xi}} \nonumber \\
&=& K \sum_{\xi=1}^4 \sigma_{\xi} \left( \v{\xi} \cdot \v{E} \right)^2 \cdot \v{\xi} \quad , \label{eq:Polarization 1} \\
\mbox{with} \; &K& = -\frac{3}{v_c} \left( \frac{e\gamma}{E_g} \right)^3  (1-S^2)\; .\nonumber
\end{eqnarray}
Using the corresponding values for Si, taking $\gamma=1.4$ \cite{HarrisonBook1989} and $S=0.5$ \cite{Harrison1974}, we have $K = -1.18\times 10^{29}$C$^3$m$^{-3}$eV$^{-3}$.

For a bond length $d$, the bond vectors $\v{\xi}$ in the crystal coordinates $\{\versor{1}, \versor{2}, \versor{3}\}=\{[100], [010], [001]\}$  are given by
\begin{eqnarray}\label{eq:bonds}
\v{\xi_1} = \frac{d}{\sqrt{3}}(1,1,1)\quad &,& \quad \v{\xi_2} = \frac{d}{\sqrt{3}}(1,-1,-1) \\
\v{\xi_3} = \frac{d}{\sqrt{3}}(-1,1,-1) \quad &,& \quad\v{\xi_4} = \frac{d}{\sqrt{3}}(-1,-1,1)
\end{eqnarray}
and replacing these coordinates in equation \ref{eq:Polarization 1}, we can write the final components of the 2\textsuperscript{nd} order nonlinear polarization in the crystal coordinates as
\begin{eqnarray}
P_x &=& \frac{K d^3}{3\sqrt{3}}
\left[
\left(\sigma_1+\sigma_2-\sigma_3-\sigma_4 \right) \left(E_x^2+E_y^2+E_z^2\right) \right. \nonumber \\
&&+ 2\left(\sigma_1-\sigma_2+\sigma_3-\sigma_4 \right) E_xE_y \nonumber \\
&&+ 2\left(\sigma_1-\sigma_2-\sigma_3+\sigma_4 \right) E_xE_z \nonumber \\
&&+ \left. 2\left(\sigma_1+\sigma_2+\sigma_3+\sigma_4 \right) E_yE_z \right] \label{eq:Px} \\
 \nonumber \\
%
P_y &=& \frac{K d^3}{3\sqrt{3}}
\left[
\left(\sigma_1-\sigma_2+\sigma_3-\sigma_4 \right) \left(E_x^2+E_y^2+E_z^2\right) \right. \nonumber \\
&&+ 2\left(\sigma_1+\sigma_2-\sigma_3-\sigma_4 \right) E_xE_y \nonumber \\
&&+ 2\left(\sigma_1-\sigma_2-\sigma_3+\sigma_4 \right) E_yE_z \nonumber \\
&&+ \left. 2\left(\sigma_1+\sigma_2+\sigma_3+\sigma_4 \right) E_xE_z \right]\label{eq:Py} \\
\nonumber \\
%
P_z &=& \frac{K d^3}{3\sqrt{3}}
\left[
\left(\sigma_1-\sigma_2-\sigma_3+\sigma_4 \right) \left(E_x^2+E_y^2+E_z^2\right) \right. \nonumber \\
&&+ 2\left(\sigma_1+\sigma_2-\sigma_3-\sigma_4 \right) E_xE_z \nonumber \\
&&+ 2\left(\sigma_1-\sigma_2+\sigma_3-\sigma_4 \right) E_yE_z \nonumber \\
&&+ \left. 2\left(\sigma_1+\sigma_2+\sigma_3+\sigma_4 \right) E_xE_y \right]\label{eq:Pz}
\end{eqnarray}

The previous set of equations, together with the definition
\begin{equation}\label{eq:chitwo}
P^{(2)}_i = 2\epsilon_0 \chi^{(2)}_{ijk} E_j E_k
\end{equation}
determines every and each component of the $\chi^{(2)}$ tensor in the crystal coordinates in terms of the polarity of each bond $\sigma_{\xi}$ in the unit cell, as represented in Fig. \ref{fig:UnitCellPol} b). This polarity, depends on the sum of the strain gradients projected on the direction of each bond, as shown by equation \ref{eq:XI}, leading to a different polarity of each bond in the unit cell. Therefore, in general, every component of \chitwo will be non-zero, contrasting with the case of a zync-blend crystal where each bond has the same polarity $\sigma$ and in which case equations \ref{eq:Px}, \ref{eq:Py} and \ref{eq:Pz} lead to the well known fact that $\chi^{(2)}_{xyz}$ is the only nonvanishing component.

The explicit calculation of the \chitwo components in terms of the strain gradients, requires an explicit expansion of the sum defining $\sigma_{\xi}$ in equation \ref{eq:polarity final}. Despite being always possible to do it in terms of the unknown parameters $\alpha$ and $\beta$, we will not write it explicitly for a general case, not only because it turns out to be a very big expression, but also it depends on the lab coordinate system relevant for that particular situation. Therefore, we will apply the previous formulations of \chitwo in strained silicon to a relevant device where we can actually analyse how the strain enables its \chitwo effects.
\section{Evaluation of the proposed model}\label{sec:evaluation}

To fully validate the model presented in the previous sections, an experimental confirmation would be required. However, as already briefly mentioned in the introduction, very recently it has been shown that the experimental data available in the literature on \chitwo phenomena in strained silicon, has a strong contribution from free carriers effects\cite{Azadeh2015,Sharma2015,Schriever2015}. This parasitic response masks the real value of strain induced \chitwo in silicon, resulting in erroneous experimental data.

As a result, most of the quantitative values of \chitwo in strained silicon presented in the literature \cite{Chmielak2011a,Chmielak2013,Damas2014,Puckett2014} have been discredited and no reliable data is available to confidently compare with the results from our model. Nevertheless, there are some properties of strain-induced \chitwo in silicon that are sill known to be valid and we will apply our model to a practical device and take conclusions regarding those characteristics.

\begin{figure}
\includegraphics[width=0.45\textwidth]{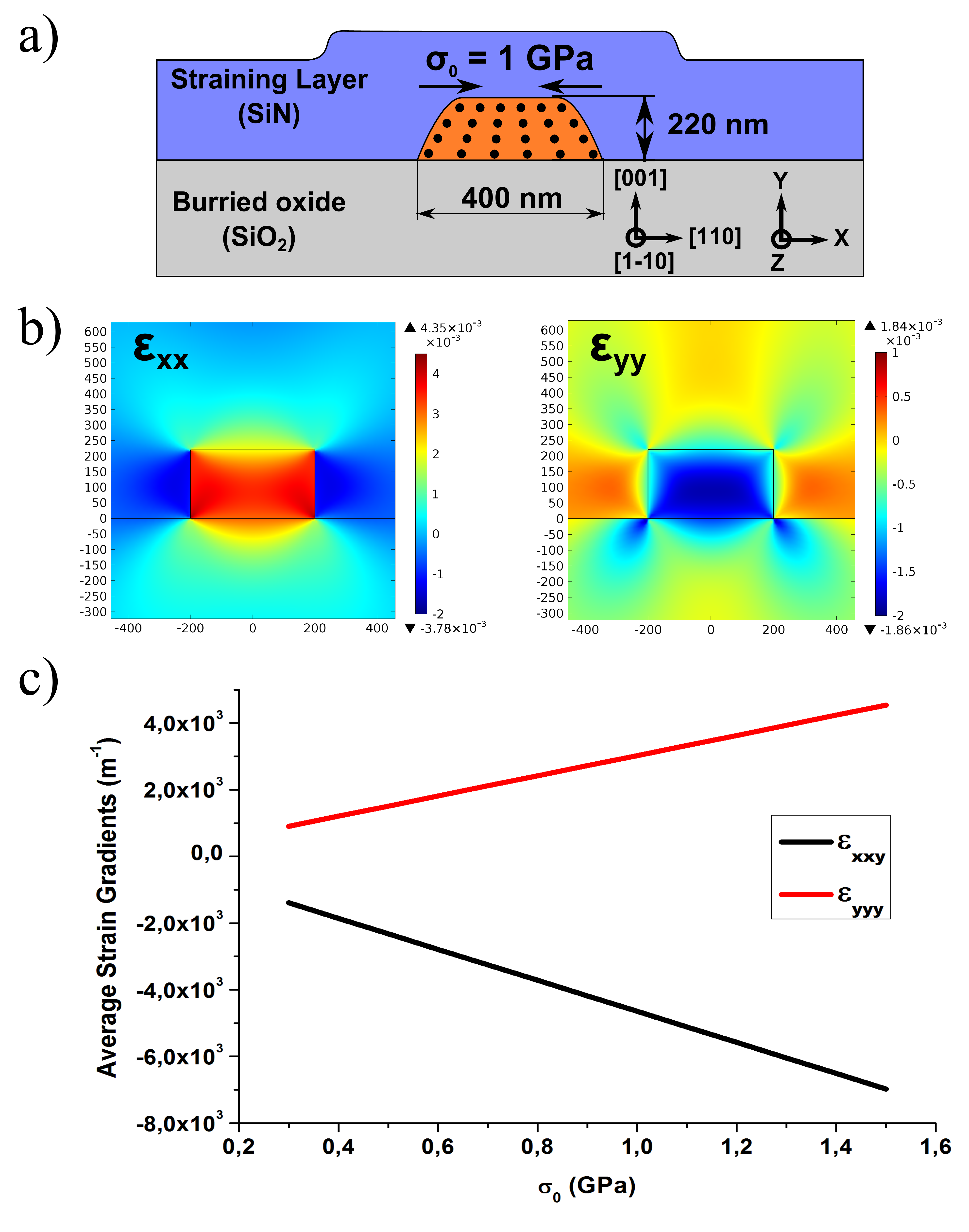} 
\caption{a) Cross section of the strained silicon device under consideration. b) Principal strain components $\varepsilon_{xx}(x,y)$ and $\varepsilon_{yy}(x,y)$ generated by the strain overlayer with $\sigma_0 = 1$ GPa. c) Linear relation between the average strain gradients components $\varepsilon_{xxy} = \partial \varepsilon_{xx}/\partial y$ and $\varepsilon_{yyy} = \partial \varepsilon_{yy}/\partial y$ in the waveguide and the initial stress $\sigma_0$.
\label{fig:waveguide}}
\end{figure}

We will evaluate the validity of our theory by applying it to the waveguide structure shown in Fig. \ref{fig:waveguide} a), which is the structure usually used in strained silicon devices towards Pockels effect modulation \cite{Chmielak2011a, Chmielak2013, Damas2014}. The waveguide coordinates $\{\versor{x},\versor{y},\versor{z}\}$ are then given in terms of the crystal coordinates $\{\versor{1},\versor{2},\versor{3}\}$ by
\begin{eqnarray*}
\versor{x} &=& \frac{1}{\sqrt{2}}\left(\versor{1}+\versor{2}\right) \\
\versor{y} &=& \versor{3} \\
\versor{z} &=& \frac{1}{\sqrt{2}}\left(\versor{1}-\versor{2}\right)
\end{eqnarray*}

The straining layer placed on top of the waveguide has an initial stress $\sigma_0$, which will induce a strain field $\strain(\v{r})$ in the waveguide. Since the waveguide extends over the $z$ direction, the $z$ strain/strain gradients components can be neglected. In addition, our simulations show that the cross strain components $\varepsilon_{ij},\;i\neq j$ inside the Si waveguide are much smaller than the principal ones $\varepsilon_{ii}$, so we will neglect their contribution in our analysis. Therefore, by defining the strain gradient components by
\begin{equation*}
\varepsilon_{ijk} \equiv \frac{\partial \varepsilon_{ij}}{\partial x_k}
\end{equation*}
we are only left with the components $\varepsilon_{xxx}$, $\varepsilon_{yyx}$, $\varepsilon_{xxy}$ and $\varepsilon_{yyy}$.

The strain induced bond polarities (eq. \ref{eq:polarity final}) in the waveguide coordinates, are calculated after rewriting the bond vector coordinates (Eq. \ref{eq:bonds}) in the $\{\versor{x},\versor{y},\versor{z}\}$ lab basis. The \chitwo components in the lab coordinates can be extracted after replacing the polarities $\sigma_\xi$ into equations \ref{eq:Px}, \ref{eq:Py} and \ref{eq:Pz} for the macroscopic polarization, in the lab coordinates. After these calculations are performed, the relevant \chitwo components for this device become:
\begin{eqnarray}
\chi^{(2)}_{xxx} &=& \frac{2 d^6 K}{27 \epsilon_0} (2 (\beta -\alpha ) \varepsilon_{xxx}+ (3 \beta-\alpha)\varepsilon_{yyx}) \label{eq:chitwo_xxx}\\
\chi^{(2)}_{yyx} &=& \chi^{(2)}_{xxx} \label{eq:chitwo_yyx}\\
\chi^{(2)}_{xxy} &=& \frac{2 d^6 K}{27 \epsilon_0} (2(\beta -\alpha ) \varepsilon_{xxy}+ (3 \beta-\alpha)\varepsilon_{yyy})  \label{eq:chitwo_xxy}\\
\chi^{(2)}_{yyy} &=& \frac{d^6 K}{27 \epsilon_0} (3 \beta - \alpha )(\varepsilon_{xxy}+\varepsilon_{yyy}) \label{eq:chitwo_yyy}
\end{eqnarray}

This set of equations gives us all the required information about the \chitwo tensor in any point of space in terms of the strain gradients in that same point. We can now compare this result with the main claims on strain-induced \chitwo in silicon.

We start by noticing that the previous equations have the form
\begin{equation}\label{eq:chitwo_strain}
\chi^{(2)}_{ijk} = \sum_{lmn} \Gamma^{ijk}_{lmn} \varepsilon_{lmn}
\end{equation}
which is a linear combination of strain gradients, as previously suggested in \cite{Manganelli2015,Damas2013}. Moreover, from our model the coefficients $\Gamma^{ijk}_{lmn}$ are known and depend only on $\alpha$ and $\beta$. For instance, from equation \ref{eq:chitwo_xxy} we extract
\begin{equation}
\Gamma^{xxy}_{xxy} = \frac{4 d^6 K}{27 \epsilon_0} (\beta -\alpha ) \quad, \quad \Gamma^{xxy}_{yyy} = \frac{2 d^6 K}{27 \epsilon_0}(3 \beta-\alpha)
\end{equation}
and this can be done to any coefficient $\Gamma^{ijk}_{lmn}$, always in terms of only $\alpha$ and $\beta$. This is in line with the claims that \chitwo should be proportional to strain gradients and not to strain itself, as it has been suggested in many publications in the past years \cite{Cazzanelli2012,Chmielak2013,Puckett2014,Damas2014,Manganelli2015}; more importantly it gives the exact value of the weight of each strain gradient direction for the desired \chitwo component. 

Other known experimental fact of strained silicon is that \chitwo has a linear relationship with the initial stress $\sigma_0$ in the straining layer \cite{Huang1994,Schriever2010,Schriever2012}.

In Fig. \ref{fig:waveguide} c), we see the simulation of the average $\varepsilon_{xxy}$ and $\varepsilon_{yyy}$ in the waveguide, for different values of $\sigma_0$ and it is clear the linear relationship between these 2 quantities. This is true for any $\varepsilon_{ijk}$ component. 
Since \chitwo is linear with $\varepsilon_{ijk}$, it is straightforward to conclude that, regardless of the values of $\alpha$ and $\beta$, our model predicts
\begin{equation}
\chi^{(2)} \propto \sigma_0\;,
\end{equation}
which is coherent with the experimental data in \cite{Schriever2010}.

\subsection{Estimation of the order of magnitude of \chitwo}\label{sec:numerical}

As it can be seen from eqs. \ref{eq:chitwo_xxx} - \ref{eq:chitwo_yyy}, \chitwo depends entirely on the parameters $\alpha$ and $\beta$, defined in Eq. \ref{eq:alphaBeta}. To determine these two parameters, the best approach would be to fit the experimental data to the proposed model and extract the values of $\alpha$ and $\beta$ that give the best fit.

However, as already mentioned, all the quantitative values of strain induced \chitwo in silicon published in the literature, in particular those in \cite{Chmielak2011a, Chmielak2013, Puckett2014, Damas2014} have strong parasitic contributions from carriers \cite{Azadeh2015,Sharma2015,Schriever2015}. Therefore, no reliable numerical data for \chitwo in strained silicon is available right now to allow for a confident fitting of $\alpha$ or $\beta$ and their evaluation must be done by approaching the definition in eq. \ref{eq:alphaBeta}.

\begin{figure}
\includegraphics[width=0.45\textwidth]{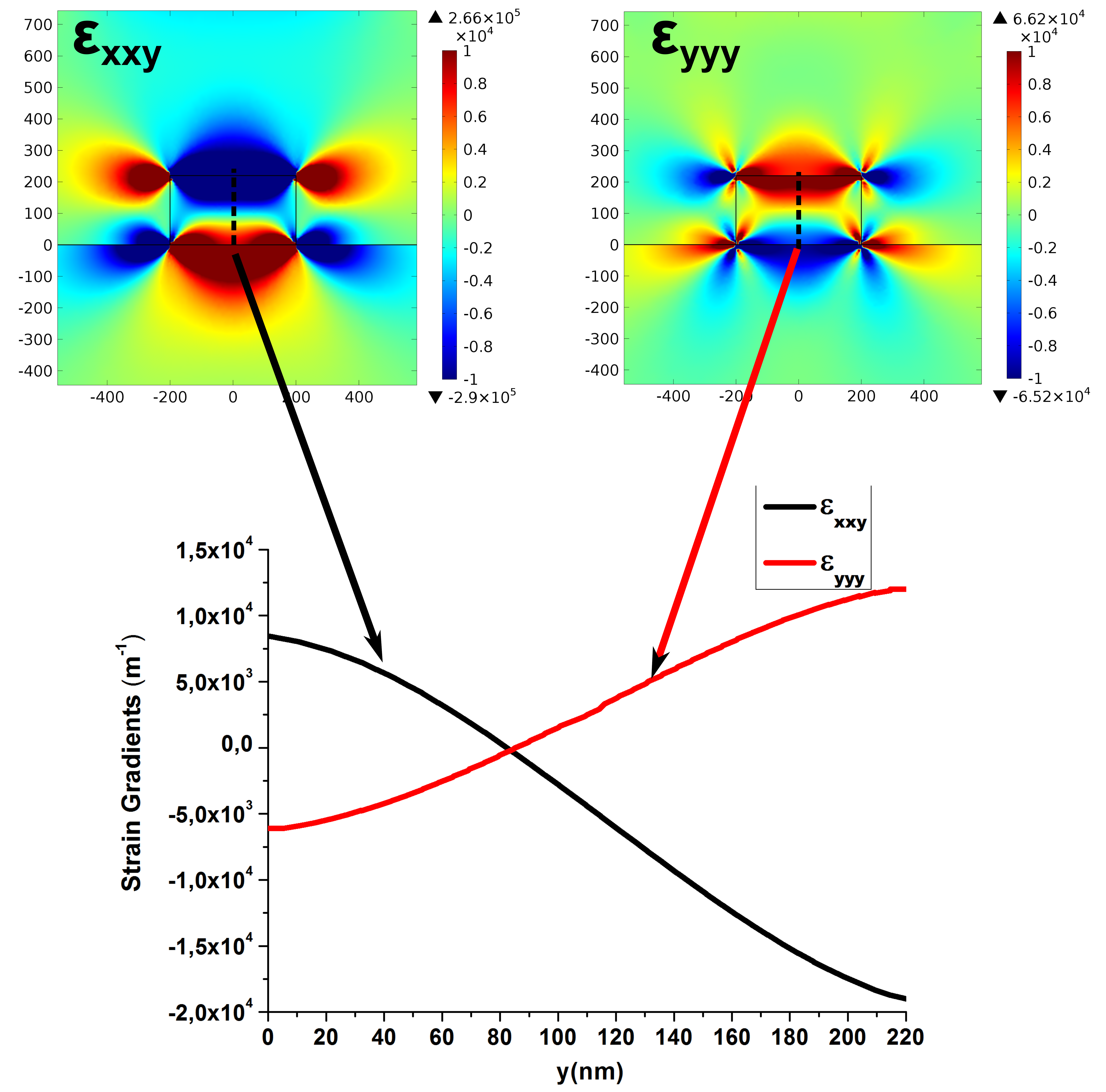} 
\caption{Strain gradients components $\varepsilon_{xxy}(x,y)$ and $\varepsilon_{yyy}(x,y)$ distribution with $\sigma_0 = 1$ GPa and the corresponding values over the vertical dashed black line throughout the center of the waveguide.
\label{fig:stressDistribution}}
\end{figure}

The evaluation of the integral in eq. \ref{eq:theta def} is not a simple task, not only because it is a difficult integral to evaluate, but mainly because the real form of the silicon crystal potential $V(\v{r})$ must be entirely known. The potential $V(\v{r})$ is recognized to be difficult to know exactly \cite{Witzens2014}, so any result deduced from $V(\v{r})$ will always have associated errors. Newertheless, the order of magnitude of the predicted values, must present a considerable level of agreement with the most recent experimental results on strained silicon. Therefore, we will focus on determining the order of magnitude of \chitwo in eqs. \ref{eq:chitwo_xxx}-\ref{eq:chitwo_yyy}.

In order to evaluate the order of magnitude of $\v{\theta}^{\xi}_i$, we must simplify the integral in eq. \ref{eq:theta def}. To do that, we will make the approximation ${\lVert \v{r}-\v{R}_{A_i} \rVert} \sim {\lVert \v{R}_{A} - \v{R}_{A_i} \rVert} = d$, which basically means that the hybrid wavefunction $\hybrid{A}(\v{r})$ is considered to be strong only close to the original atom. Despite this is not entirely true because it extends along its bond, this simplification should not change considerably the order of magnitude of the integral of eq. \ref{eq:theta def}. In that case, eq. \ref{eq:theta2} becomes
\begin{eqnarray}
\v{\theta}^{\xi}_i &\sim& \left.\frac{\partial V}{\partial r}\right\vert_{d} \frac{1}{d}\int |h_{\xi}^A(\v{r})|^2 \cdot \left(\v{r}-\v{R}_{A_i}\right) dV  \\
 &\sim&  \frac{1-\gamma}{2d} \cdot \left.\frac{\partial V}{\partial r}\right\vert_{d}\v{\xi} - \frac{1}{d}\left.\frac{\partial V}{\partial r}\right\vert_{d}\v{\xi}_i
\end{eqnarray}
The previous equation gives us values for $\alpha$ and $\beta$ which are merely approximations, but should be in the same order of magnitude of the real ones:
\begin{equation}\label{eq:explicit alpha/beta}
\alpha \sim \frac{1-\gamma}{2d} \cdot \left.\frac{\partial V}{\partial r}\right\vert_{d} \quad , \quad \beta \sim - \frac{1}{d}\left.\frac{\partial V}{\partial r}\right\vert_{d}
\end{equation}

As already mentioned, the determination of the real Si crystal potential $V(\v{r})$ is a very complex problem, which has been studied for many years \cite{Appelbaum1973,Anderson1969,Chelikowsky1991,Wendel1978}. Because of its complexity, in this work we will only compare two simple crystal potentials to extract some numerical information: the Coulomb potential generated by a Si$^{4+}$ ion
\begin{equation}
V_c(\v{r}) = - \kappa \frac{4}{r}
\end{equation}
and the Phillips potential for Si \cite{Phillips1959}
\begin{equation}
V_{P}(\v{r}) = -\kappa  \left(\frac{2 (A-Z) e^{-\lambda  r}}{r}+\frac{2 Z}{r}\right)
\end{equation}
with $\kappa = e^2/(4\pi \epsilon_0) = 2.3\times10^{-28}$ kg m$^3$/s$^2$ and the parameters $A=13.1$, $Z=3.1$ and $\lambda = 1.64\times10^{-10}$ m$^{-1}$.

Using eq. \ref{eq:explicit alpha/beta} for both of these potentials, taking $d=0.235$ nm and focusing only the $\chi_{xxy}^{(2)}$ component (which is the one that has been more strongly studied in the literature \cite{Jacobsen2006, Chmielak2011a,Chmielak2013, Puckett2014, Damas2014}), we get for the Coulomb and Phillips potentials, respectively (in S.I. units):
\begin{eqnarray}
\chi_{xxy_{c}}^{(2)} &\sim& 8.0\times 10^{-17} \varepsilon_{yyy} + 4.6\times 10^{-17}\varepsilon_{xxy} \\
\chi_{xxy_{P}}^{(2)} &\sim& 5.2\times 10^{-16} \varepsilon_{yyy} + 3.0\times 10^{-16}\varepsilon_{xxy} \; .
\end{eqnarray}

Now, as shown in Fig. \ref{fig:stressDistribution}, the typical order of magnitude of the strain gradients on the edges of the waveguide (where the applied electric field is stronger \cite{Sharma2015}) is $ \sim 10^4\mbox{ m}^{-1}$, leading to:
\begin{equation} \label{eq:numerical}
\chi_{xxy_{c}}^{(2)} \sim 1\,\mbox{pm/V}\quad , \quad \chi_{xxy_{P}}^{(2)} \sim 8\,\mbox{pm/V}
\end{equation}

The strong dependence of \chitwo on the choice of potential $V(\v{r})$ is clear from the results above. It means that any difference between predicted and experimental results can be attributed to a bad choice of the potential $V(\v{r})$ used to describe the crystal. This problem can only be overcome by fitting $\alpha$ and $\beta$ to available experimental data. Any other way of obtaining these two parameters, will inevitably have errors associated because the used potential $V(\v{r})$ will always be an approximation to the real and much more complex potential felt by the electrons in a real crystal. 

On the other hand, it is not straightforward which published experimental values we should compare to. It is now widely accepted that previously reported values of \chitwo on the order of magnitude of $100$ pm/V as the ones published in \cite{Chmielak2011a,Chmielak2013,Damas2014} were wrongly interpreted and they are mainly due to free carriers effects and not to strain. In fact, latest results, which account for carriers effects, suggest values of \chitwo much lower than these and their order of magnitude should be around $10$ pm/V \cite{Schriever2015} or even as low as $1$ pm/V \cite{Borghi2015b}. Moreover, even these lower values of \chitwo could have been erroneously estimated as latest publications suggest that the electric field applied to the waveguide to induce Pockels effect modulation (the method used to experimentally obtain these values) is not homogeneous through the waveguide, but strongly modified by carriers effects \cite{Sharma2015}, which does not seem to have been taken into account in these publications.

Nevertheless, we see that the values presented by our model in equation \ref{eq:numerical}, even though we used very simplified potentials $V(\v{r})$, are very close, in order of magnitude, to the latest experimental results available on strain induced \chitwo .
%
%
\section{Conclusion}
In this work we develop and present an atomistic model, based on the \textit{bond orbital model} to describe the second order nonlinear effects generated by strain in the silicon crystal. This model gives a spatial quantitative and well defined relation between the \chitwo tensor and the strain tensor \strain .

We have shown that \chitwo is proportional to a weigthed sum of strain gradient components, as suggested by many publications. The weighting coefficients depend only on 2 coefficients, $\alpha$ and $\beta$, which can be theoretically estimated, but should be experimentally determined to fully validate this model.

By applying this model to a specific geometry, with the characteristics of the main devices used in strained silicon photonics for Pockels effect modulation, we were able to show agreement of our model with the known properties of strain induced \chitwo in silicon. Furthermore, we estimated the order of magnitude of a component of \chitwo calculated using our model and its values (between 1 pm/V and 8 pm/V) showed a considerable agreement with the latest published experimental results. Nevertheless, this value can be strongly improved once reliable experimental data is available for a confident fit of the numerical predictions of this model.

We consider that the presented model is of extreme relevance for the study of nonlinear effects in strained silicon photonics. With the relation between \chitwo and \strain that we developed in this paper, the optimization of strained silicon devices is finally possible. The strain distribution in the crystal can be engineered to maximize the most relevant \chitwo components for the desired device and this opens a whole new route towards the improvement of nonlinear effects in strained silicon, bringing us closer to high performance devices based on this kind of effects.

\acknowledgments
Authors would like to thank Xavier Le Roux from IEF and Fr\'ed\'eric Boeuf from STMicroelectronics (Crolles, France) for fruitful discussions. The authors also acknowledge STMicroelectronics for the financial support of the P. Damas' scholarship. This project has received funding from the European Research Council (ERC) under the European Union's Horizon 2020 research and innovation program (ERC POPSTAR - grant agreement N\textsuperscript{o} 647342).

%
\bibliographystyle{apsrev4-1}	
\bibliography{bibliography}		

\begin{thebibliography}{42}%
\makeatletter
\providecommand \@ifxundefined [1]{%
 \@ifx{#1\undefined}
}%
\providecommand \@ifnum [1]{%
 \ifnum #1\expandafter \@firstoftwo
 \else \expandafter \@secondoftwo
 \fi
}%
\providecommand \@ifx [1]{%
 \ifx #1\expandafter \@firstoftwo
 \else \expandafter \@secondoftwo
 \fi
}%
\providecommand \natexlab [1]{#1}%
\providecommand \enquote  [1]{``#1''}%
\providecommand \bibnamefont  [1]{#1}%
\providecommand \bibfnamefont [1]{#1}%
\providecommand \citenamefont [1]{#1}%
\providecommand \href@noop [0]{\@secondoftwo}%
\providecommand \href [0]{\begingroup \@sanitize@url \@href}%
\providecommand \@href[1]{\@@startlink{#1}\@@href}%
\providecommand \@@href[1]{\endgroup#1\@@endlink}%
\providecommand \@sanitize@url [0]{\catcode `\\12\catcode `\$12\catcode
  `\&12\catcode `\#12\catcode `\^12\catcode `\_12\catcode `\%12\relax}%
\providecommand \@@startlink[1]{}%
\providecommand \@@endlink[0]{}%
\providecommand \url  [0]{\begingroup\@sanitize@url \@url }%
\providecommand \@url [1]{\endgroup\@href {#1}{\urlprefix }}%
\providecommand \urlprefix  [0]{URL }%
\providecommand \Eprint [0]{\href }%
\providecommand \doibase [0]{http://dx.doi.org/}%
\providecommand \selectlanguage [0]{\@gobble}%
\providecommand \bibinfo  [0]{\@secondoftwo}%
\providecommand \bibfield  [0]{\@secondoftwo}%
\providecommand \translation [1]{[#1]}%
\providecommand \BibitemOpen [0]{}%
\providecommand \bibitemStop [0]{}%
\providecommand \bibitemNoStop [0]{.\EOS\space}%
\providecommand \EOS [0]{\spacefactor3000\relax}%
\providecommand \BibitemShut  [1]{\csname bibitem#1\endcsname}%
\let\auto@bib@innerbib\@empty
\bibitem [{\citenamefont {Fedeli}\ \emph {et~al.}(2008)\citenamefont {Fedeli},
  \citenamefont {Cioccio}, \citenamefont {Marris-Morini}, \citenamefont
  {Vivien}, \citenamefont {Orobtchouk}, \citenamefont {Rojo-Romeo},
  \citenamefont {Seassal},\ and\ \citenamefont {Mandorlo}}]{Fedeli2008}%
  \BibitemOpen
  \bibfield  {author} {\bibinfo {author} {\bibfnamefont {J.~M.}\ \bibnamefont
  {Fedeli}}, \bibinfo {author} {\bibfnamefont {L.~D.}\ \bibnamefont {Cioccio}},
  \bibinfo {author} {\bibfnamefont {D.}~\bibnamefont {Marris-Morini}}, \bibinfo
  {author} {\bibfnamefont {L.}~\bibnamefont {Vivien}}, \bibinfo {author}
  {\bibfnamefont {R.}~\bibnamefont {Orobtchouk}}, \bibinfo {author}
  {\bibfnamefont {P.}~\bibnamefont {Rojo-Romeo}}, \bibinfo {author}
  {\bibfnamefont {C.}~\bibnamefont {Seassal}}, \ and\ \bibinfo {author}
  {\bibfnamefont {F.}~\bibnamefont {Mandorlo}},\ }\href@noop {} {\bibfield
  {journal} {\bibinfo  {journal} {Adv. Optical Technol., Special issue on
  "Silicon Photonics"}\ } (\bibinfo {year} {2008})}\BibitemShut {NoStop}%
\bibitem [{\citenamefont {Leuthold}\ \emph {et~al.}(2010)\citenamefont
  {Leuthold}, \citenamefont {Koos},\ and\ \citenamefont
  {Freude}}]{Leuthold2010}%
  \BibitemOpen
  \bibfield  {author} {\bibinfo {author} {\bibfnamefont {J.}~\bibnamefont
  {Leuthold}}, \bibinfo {author} {\bibfnamefont {C.}~\bibnamefont {Koos}}, \
  and\ \bibinfo {author} {\bibfnamefont {W.}~\bibnamefont {Freude}},\ }\href
  {\doibase 10.1038/nphoton.2010.185} {\bibfield  {journal} {\bibinfo
  {journal} {Nature Photonics}\ }\textbf {\bibinfo {volume} {4}},\ \bibinfo
  {pages} {535} (\bibinfo {year} {2010})}\BibitemShut {NoStop}%
\bibitem [{\citenamefont {Reed}\ \emph {et~al.}(2010)\citenamefont {Reed},
  \citenamefont {Mashanovich}, \citenamefont {Gardes},\ and\ \citenamefont
  {Thomson}}]{Reed2010}%
  \BibitemOpen
  \bibfield  {author} {\bibinfo {author} {\bibfnamefont {G.~T.}\ \bibnamefont
  {Reed}}, \bibinfo {author} {\bibfnamefont {G.}~\bibnamefont {Mashanovich}},
  \bibinfo {author} {\bibfnamefont {F.~Y.}\ \bibnamefont {Gardes}}, \ and\
  \bibinfo {author} {\bibfnamefont {D.~J.}\ \bibnamefont {Thomson}},\ }\href
  {\doibase 10.1038/nphoton.2010.179} {\bibfield  {journal} {\bibinfo
  {journal} {Nature Photonics}\ }\textbf {\bibinfo {volume} {4}},\ \bibinfo
  {pages} {518} (\bibinfo {year} {2010})}\BibitemShut {NoStop}%
\bibitem [{\citenamefont {Jacobsen}\ \emph {et~al.}(2006)\citenamefont
  {Jacobsen}, \citenamefont {Andersen}, \citenamefont {Borel}, \citenamefont
  {Fage-Pedersen}, \citenamefont {Frandsen}, \citenamefont {Hansen},
  \citenamefont {Kristensen}, \citenamefont {Lavrinenko}, \citenamefont
  {Moulin}, \citenamefont {Ou}, \citenamefont {Peucheret}, \citenamefont
  {Zsigri},\ and\ \citenamefont {Bjarklev}}]{Jacobsen2006}%
  \BibitemOpen
  \bibfield  {author} {\bibinfo {author} {\bibfnamefont {R.~S.}\ \bibnamefont
  {Jacobsen}}, \bibinfo {author} {\bibfnamefont {K.~N.}\ \bibnamefont
  {Andersen}}, \bibinfo {author} {\bibfnamefont {P.~I.}\ \bibnamefont {Borel}},
  \bibinfo {author} {\bibfnamefont {J.}~\bibnamefont {Fage-Pedersen}}, \bibinfo
  {author} {\bibfnamefont {L.~H.}\ \bibnamefont {Frandsen}}, \bibinfo {author}
  {\bibfnamefont {O.}~\bibnamefont {Hansen}}, \bibinfo {author} {\bibfnamefont
  {M.}~\bibnamefont {Kristensen}}, \bibinfo {author} {\bibfnamefont {A.~V.}\
  \bibnamefont {Lavrinenko}}, \bibinfo {author} {\bibfnamefont
  {G.}~\bibnamefont {Moulin}}, \bibinfo {author} {\bibfnamefont
  {H.}~\bibnamefont {Ou}}, \bibinfo {author} {\bibfnamefont {C.}~\bibnamefont
  {Peucheret}}, \bibinfo {author} {\bibfnamefont {B.}~\bibnamefont {Zsigri}}, \
  and\ \bibinfo {author} {\bibfnamefont {A.}~\bibnamefont {Bjarklev}},\ }\href
  {\doibase 10.1038/nature04706} {\bibfield  {journal} {\bibinfo  {journal}
  {Nature}\ }\textbf {\bibinfo {volume} {441}},\ \bibinfo {pages} {199}
  (\bibinfo {year} {2006})}\BibitemShut {NoStop}%
\bibitem [{\citenamefont {Chmielak}\ \emph {et~al.}(2011)\citenamefont
  {Chmielak}, \citenamefont {Waldow}, \citenamefont {Matheisen}, \citenamefont
  {Ripperda}, \citenamefont {Bolten}, \citenamefont {Wahlbrink}, \citenamefont
  {Nagel}, \citenamefont {Merget},\ and\ \citenamefont {Kurz}}]{Chmielak2011a}%
  \BibitemOpen
  \bibfield  {author} {\bibinfo {author} {\bibfnamefont {B.}~\bibnamefont
  {Chmielak}}, \bibinfo {author} {\bibfnamefont {M.}~\bibnamefont {Waldow}},
  \bibinfo {author} {\bibfnamefont {C.}~\bibnamefont {Matheisen}}, \bibinfo
  {author} {\bibfnamefont {C.}~\bibnamefont {Ripperda}}, \bibinfo {author}
  {\bibfnamefont {J.}~\bibnamefont {Bolten}}, \bibinfo {author} {\bibfnamefont
  {T.}~\bibnamefont {Wahlbrink}}, \bibinfo {author} {\bibfnamefont
  {M.}~\bibnamefont {Nagel}}, \bibinfo {author} {\bibfnamefont
  {F.}~\bibnamefont {Merget}}, \ and\ \bibinfo {author} {\bibfnamefont
  {H.}~\bibnamefont {Kurz}},\ }\href {\doibase 10.1364/OE.19.017212} {\bibfield
   {journal} {\bibinfo  {journal} {Optics express}\ }\textbf {\bibinfo {volume}
  {19}},\ \bibinfo {pages} {17212} (\bibinfo {year} {2011})}\BibitemShut
  {NoStop}%
\bibitem [{\citenamefont {Chmielak}\ \emph {et~al.}(2013)\citenamefont
  {Chmielak}, \citenamefont {Matheisen}, \citenamefont {Ripperda},
  \citenamefont {Bolten}, \citenamefont {Wahlbrink}, \citenamefont {Waldow},\
  and\ \citenamefont {Kurz}}]{Chmielak2013}%
  \BibitemOpen
  \bibfield  {author} {\bibinfo {author} {\bibfnamefont {B.}~\bibnamefont
  {Chmielak}}, \bibinfo {author} {\bibfnamefont {C.}~\bibnamefont {Matheisen}},
  \bibinfo {author} {\bibfnamefont {C.}~\bibnamefont {Ripperda}}, \bibinfo
  {author} {\bibfnamefont {J.}~\bibnamefont {Bolten}}, \bibinfo {author}
  {\bibfnamefont {T.}~\bibnamefont {Wahlbrink}}, \bibinfo {author}
  {\bibfnamefont {M.}~\bibnamefont {Waldow}}, \ and\ \bibinfo {author}
  {\bibfnamefont {H.}~\bibnamefont {Kurz}},\ }\href {\doibase
  10.1364/OE.21.025324} {\bibfield  {journal} {\bibinfo  {journal} {Optics
  Express}\ }\textbf {\bibinfo {volume} {21}},\ \bibinfo {pages} {25324}
  (\bibinfo {year} {2013})}\BibitemShut {NoStop}%
\bibitem [{\citenamefont {Puckett}\ \emph {et~al.}(2014)\citenamefont
  {Puckett}, \citenamefont {Smalley}, \citenamefont {Abashin}, \citenamefont
  {Grieco},\ and\ \citenamefont {Fainman}}]{Puckett2014}%
  \BibitemOpen
  \bibfield  {author} {\bibinfo {author} {\bibfnamefont {M.~W.}\ \bibnamefont
  {Puckett}}, \bibinfo {author} {\bibfnamefont {J.~S.~T.}\ \bibnamefont
  {Smalley}}, \bibinfo {author} {\bibfnamefont {M.}~\bibnamefont {Abashin}},
  \bibinfo {author} {\bibfnamefont {A.}~\bibnamefont {Grieco}}, \ and\ \bibinfo
  {author} {\bibfnamefont {Y.}~\bibnamefont {Fainman}},\ }\href
  {http://www.ncbi.nlm.nih.gov/pubmed/24690871} {\bibfield  {journal} {\bibinfo
   {journal} {Optics letters}\ }\textbf {\bibinfo {volume} {39}},\ \bibinfo
  {pages} {1693} (\bibinfo {year} {2014})}\BibitemShut {NoStop}%
\bibitem [{\citenamefont {Damas}\ \emph {et~al.}(2014)\citenamefont {Damas},
  \citenamefont {{Le Roux}}, \citenamefont {{Le Bourdais}}, \citenamefont
  {Cassan}, \citenamefont {Marris-Morini}, \citenamefont {Izard}, \citenamefont
  {Maroutian}, \citenamefont {Lecoeur},\ and\ \citenamefont
  {Vivien}}]{Damas2014}%
  \BibitemOpen
  \bibfield  {author} {\bibinfo {author} {\bibfnamefont {P.}~\bibnamefont
  {Damas}}, \bibinfo {author} {\bibfnamefont {X.}~\bibnamefont {{Le Roux}}},
  \bibinfo {author} {\bibfnamefont {D.}~\bibnamefont {{Le Bourdais}}}, \bibinfo
  {author} {\bibfnamefont {E.}~\bibnamefont {Cassan}}, \bibinfo {author}
  {\bibfnamefont {D.}~\bibnamefont {Marris-Morini}}, \bibinfo {author}
  {\bibfnamefont {N.}~\bibnamefont {Izard}}, \bibinfo {author} {\bibfnamefont
  {T.}~\bibnamefont {Maroutian}}, \bibinfo {author} {\bibfnamefont
  {P.}~\bibnamefont {Lecoeur}}, \ and\ \bibinfo {author} {\bibfnamefont
  {L.}~\bibnamefont {Vivien}},\ }\href {\doibase 10.1364/OE.22.022095}
  {\bibfield  {journal} {\bibinfo  {journal} {Optics Express}\ }\textbf
  {\bibinfo {volume} {22}},\ \bibinfo {pages} {22095} (\bibinfo {year}
  {2014})}\BibitemShut {NoStop}%
\bibitem [{\citenamefont {Cazzanelli}\ \emph {et~al.}(2012)\citenamefont
  {Cazzanelli}, \citenamefont {Bianco}, \citenamefont {Borga}, \citenamefont
  {Pucker}, \citenamefont {Ghulinyan}, \citenamefont {Degoli}, \citenamefont
  {Luppi}, \citenamefont {V\'{e}niard}, \citenamefont {Ossicini}, \citenamefont
  {Modotto}, \citenamefont {Wabnitz}, \citenamefont {Pierobon},\ and\
  \citenamefont {Pavesi}}]{Cazzanelli2012}%
  \BibitemOpen
  \bibfield  {author} {\bibinfo {author} {\bibfnamefont {M.}~\bibnamefont
  {Cazzanelli}}, \bibinfo {author} {\bibfnamefont {F.}~\bibnamefont {Bianco}},
  \bibinfo {author} {\bibfnamefont {E.}~\bibnamefont {Borga}}, \bibinfo
  {author} {\bibfnamefont {G.}~\bibnamefont {Pucker}}, \bibinfo {author}
  {\bibfnamefont {M.}~\bibnamefont {Ghulinyan}}, \bibinfo {author}
  {\bibfnamefont {E.}~\bibnamefont {Degoli}}, \bibinfo {author} {\bibfnamefont
  {E.}~\bibnamefont {Luppi}}, \bibinfo {author} {\bibfnamefont
  {V.}~\bibnamefont {V\'{e}niard}}, \bibinfo {author} {\bibfnamefont
  {S.}~\bibnamefont {Ossicini}}, \bibinfo {author} {\bibfnamefont
  {D.}~\bibnamefont {Modotto}}, \bibinfo {author} {\bibfnamefont
  {S.}~\bibnamefont {Wabnitz}}, \bibinfo {author} {\bibfnamefont
  {R.}~\bibnamefont {Pierobon}}, \ and\ \bibinfo {author} {\bibfnamefont
  {L.}~\bibnamefont {Pavesi}},\ }\href {\doibase 10.1038/nmat3200} {\bibfield
  {journal} {\bibinfo  {journal} {Nature materials}\ }\textbf {\bibinfo
  {volume} {11}},\ \bibinfo {pages} {148} (\bibinfo {year} {2012})}\BibitemShut
  {NoStop}%
\bibitem [{\citenamefont {Schriever}\ \emph {et~al.}(2012)\citenamefont
  {Schriever}, \citenamefont {Bohley}, \citenamefont {Schilling},\ and\
  \citenamefont {Wehrspohn}}]{Schriever2012}%
  \BibitemOpen
  \bibfield  {author} {\bibinfo {author} {\bibfnamefont {C.}~\bibnamefont
  {Schriever}}, \bibinfo {author} {\bibfnamefont {C.}~\bibnamefont {Bohley}},
  \bibinfo {author} {\bibfnamefont {J.}~\bibnamefont {Schilling}}, \ and\
  \bibinfo {author} {\bibfnamefont {R.~B.}\ \bibnamefont {Wehrspohn}},\ }\href
  {\doibase 10.3390/ma5050889} {\bibfield  {journal} {\bibinfo  {journal}
  {Materials}\ }\textbf {\bibinfo {volume} {5}},\ \bibinfo {pages} {889}
  (\bibinfo {year} {2012})}\BibitemShut {NoStop}%
\bibitem [{\citenamefont {Govorkov}\ \emph {et~al.}(1989)\citenamefont
  {Govorkov}, \citenamefont {Emel'yanov}, \citenamefont {Koroteev},
  \citenamefont {Petrov}, \citenamefont {Shumay}, \citenamefont {Yakovlev},\
  and\ \citenamefont {Khokhlov}}]{Govorkov1989}%
  \BibitemOpen
  \bibfield  {author} {\bibinfo {author} {\bibfnamefont {S.~V.}\ \bibnamefont
  {Govorkov}}, \bibinfo {author} {\bibfnamefont {V.~I.}\ \bibnamefont
  {Emel'yanov}}, \bibinfo {author} {\bibfnamefont {N.~I.}\ \bibnamefont
  {Koroteev}}, \bibinfo {author} {\bibfnamefont {G.~I.}\ \bibnamefont
  {Petrov}}, \bibinfo {author} {\bibfnamefont {I.~L.}\ \bibnamefont {Shumay}},
  \bibinfo {author} {\bibfnamefont {V.~V.}\ \bibnamefont {Yakovlev}}, \ and\
  \bibinfo {author} {\bibfnamefont {R.~V.}\ \bibnamefont {Khokhlov}},\ }\href
  {\doibase 10.1364/JOSAB.6.001117} {\bibfield  {journal} {\bibinfo  {journal}
  {Journal of the Optical Society of America B}\ }\textbf {\bibinfo {volume}
  {6}},\ \bibinfo {pages} {1117} (\bibinfo {year} {1989})}\BibitemShut
  {NoStop}%
\bibitem [{\citenamefont {Huang}(1994)}]{Huang1994}%
  \BibitemOpen
  \bibfield  {author} {\bibinfo {author} {\bibfnamefont {J.}~\bibnamefont
  {Huang}},\ }\href {http://jjap.jsap.jp/link?JJAP/33/3878/} {\bibfield
  {journal} {\bibinfo  {journal} {Jpn. J. Appi. Phys. Vol}\ }\textbf {\bibinfo
  {volume} {33}},\ \bibinfo {pages} {3878} (\bibinfo {year}
  {1994})}\BibitemShut {NoStop}%
\bibitem [{\citenamefont {Levine}(1973)}]{Levine1973}%
  \BibitemOpen
  \bibfield  {author} {\bibinfo {author} {\bibfnamefont {B.}~\bibnamefont
  {Levine}},\ }\href {http://prb.aps.org/abstract/PRB/v7/i6/p2600\_1}
  {\bibfield  {journal} {\bibinfo  {journal} {Physical Review B}\ }\textbf
  {\bibinfo {volume} {7}},\ \bibinfo {pages} {2600} (\bibinfo {year}
  {1973})}\BibitemShut {NoStop}%
\bibitem [{\citenamefont {Levine}(1969)}]{Levine1969}%
  \BibitemOpen
  \bibfield  {author} {\bibinfo {author} {\bibfnamefont {B.}~\bibnamefont
  {Levine}},\ }\href {http://link.aps.org/doi/10.1103/PhysRevLett.22.787}
  {\bibfield  {journal} {\bibinfo  {journal} {Physical Review Letters}\
  }\textbf {\bibinfo {volume} {22}},\ \bibinfo {pages} {787} (\bibinfo {year}
  {1969})}\BibitemShut {NoStop}%
\bibitem [{\citenamefont {Kleinman}(1962)}]{Kleinman1962a}%
  \BibitemOpen
  \bibfield  {author} {\bibinfo {author} {\bibfnamefont {D.}~\bibnamefont
  {Kleinman}},\ }\href {http://prola.aps.org/abstract/PR/v126/i6/p1977\_1}
  {\bibfield  {journal} {\bibinfo  {journal} {Physical Review}\ }\textbf
  {\bibinfo {volume} {126}},\ \bibinfo {pages} {1977} (\bibinfo {year}
  {1962})}\BibitemShut {NoStop}%
\bibitem [{\citenamefont {Harrison}\ and\ \citenamefont
  {Ciraci}(1974)}]{Harrison1974}%
  \BibitemOpen
  \bibfield  {author} {\bibinfo {author} {\bibfnamefont {W.}~\bibnamefont
  {Harrison}}\ and\ \bibinfo {author} {\bibfnamefont {S.}~\bibnamefont
  {Ciraci}},\ }\href {http://prb.aps.org/abstract/PRB/v10/i4/p1516\_1}
  {\bibfield  {journal} {\bibinfo  {journal} {Physical Review B}\ }\textbf
  {\bibinfo {volume} {10}} (\bibinfo {year} {1974})}\BibitemShut {NoStop}%
\bibitem [{\citenamefont {Aspnes}(2010)}]{Aspnes2010}%
  \BibitemOpen
  \bibfield  {author} {\bibinfo {author} {\bibfnamefont {D.~E.}\ \bibnamefont
  {Aspnes}},\ }\href {\doibase 10.1002/pssb.200983937} {\bibfield  {journal}
  {\bibinfo  {journal} {Physica Status Solidi (B)}\ }\textbf {\bibinfo {volume}
  {247}},\ \bibinfo {pages} {1873} (\bibinfo {year} {2010})}\BibitemShut
  {NoStop}%
\bibitem [{\citenamefont {Hon}\ \emph {et~al.}(2009{\natexlab{a}})\citenamefont
  {Hon}, \citenamefont {Tsia}, \citenamefont {Solli}, \citenamefont {Jalali},\
  and\ \citenamefont {Khurgin}}]{Hon2009a}%
  \BibitemOpen
  \bibfield  {author} {\bibinfo {author} {\bibfnamefont {N.~N.~K.}\
  \bibnamefont {Hon}}, \bibinfo {author} {\bibfnamefont {K.~K.~K.}\
  \bibnamefont {Tsia}}, \bibinfo {author} {\bibfnamefont {D.~R.~D.}\
  \bibnamefont {Solli}}, \bibinfo {author} {\bibfnamefont {B.}~\bibnamefont
  {Jalali}}, \ and\ \bibinfo {author} {\bibfnamefont {J.~B.}\ \bibnamefont
  {Khurgin}},\ }in\ \href {\doibase 10.1109/GROUP4.2009.5338380}
  {{\selectlanguage {english}\emph {\bibinfo {booktitle} {2009 6th IEEE
  International Conference on Group IV Photonics}}}}\ (\bibinfo  {publisher}
  {IEEE},\ \bibinfo {year} {2009})\ pp.\ \bibinfo {pages}
  {232--234}\BibitemShut {NoStop}%
\bibitem [{\citenamefont {Hon}\ \emph {et~al.}(2009{\natexlab{b}})\citenamefont
  {Hon}, \citenamefont {Tsia}, \citenamefont {Solli},\ and\ \citenamefont
  {Jalali}}]{Hon2009}%
  \BibitemOpen
  \bibfield  {author} {\bibinfo {author} {\bibfnamefont {N.~K.}\ \bibnamefont
  {Hon}}, \bibinfo {author} {\bibfnamefont {K.~K.}\ \bibnamefont {Tsia}},
  \bibinfo {author} {\bibfnamefont {D.~R.}\ \bibnamefont {Solli}}, \ and\
  \bibinfo {author} {\bibfnamefont {B.}~\bibnamefont {Jalali}},\ }\href
  {\doibase 10.1063/1.3094750} {\bibfield  {journal} {\bibinfo  {journal}
  {Applied Physics Letters}\ }\textbf {\bibinfo {volume} {94}},\ \bibinfo
  {pages} {091116} (\bibinfo {year} {2009}{\natexlab{b}})}\BibitemShut
  {NoStop}%
\bibitem [{\citenamefont {Luppi}\ \emph {et~al.}(2015)\citenamefont {Luppi},
  \citenamefont {Degoli}, \citenamefont {Bertocchi}, \citenamefont {Ossicini},\
  and\ \citenamefont {V\'{e}niard}}]{Luppi2015}%
  \BibitemOpen
  \bibfield  {author} {\bibinfo {author} {\bibfnamefont {E.}~\bibnamefont
  {Luppi}}, \bibinfo {author} {\bibfnamefont {E.}~\bibnamefont {Degoli}},
  \bibinfo {author} {\bibfnamefont {M.}~\bibnamefont {Bertocchi}}, \bibinfo
  {author} {\bibfnamefont {S.}~\bibnamefont {Ossicini}}, \ and\ \bibinfo
  {author} {\bibfnamefont {V.}~\bibnamefont {V\'{e}niard}},\ }\href {\doibase
  10.1103/PhysRevB.92.075204} {\bibfield  {journal} {\bibinfo  {journal}
  {Physical Review B}\ }\textbf {\bibinfo {volume} {92}},\ \bibinfo {pages}
  {075204} (\bibinfo {year} {2015})}\BibitemShut {NoStop}%
\bibitem [{\citenamefont {Schriever}\ \emph {et~al.}(2015)\citenamefont
  {Schriever}, \citenamefont {Bianco}, \citenamefont {Cazzanelli},
  \citenamefont {Ghulinyan}, \citenamefont {Eisenschmidt}, \citenamefont
  {de~Boor}, \citenamefont {Schmid}, \citenamefont {Heitmann}, \citenamefont
  {Pavesi},\ and\ \citenamefont {Schilling}}]{Schriever2015}%
  \BibitemOpen
  \bibfield  {author} {\bibinfo {author} {\bibfnamefont {C.}~\bibnamefont
  {Schriever}}, \bibinfo {author} {\bibfnamefont {F.}~\bibnamefont {Bianco}},
  \bibinfo {author} {\bibfnamefont {M.}~\bibnamefont {Cazzanelli}}, \bibinfo
  {author} {\bibfnamefont {M.}~\bibnamefont {Ghulinyan}}, \bibinfo {author}
  {\bibfnamefont {C.}~\bibnamefont {Eisenschmidt}}, \bibinfo {author}
  {\bibfnamefont {J.}~\bibnamefont {de~Boor}}, \bibinfo {author} {\bibfnamefont
  {A.}~\bibnamefont {Schmid}}, \bibinfo {author} {\bibfnamefont
  {J.}~\bibnamefont {Heitmann}}, \bibinfo {author} {\bibfnamefont
  {L.}~\bibnamefont {Pavesi}}, \ and\ \bibinfo {author} {\bibfnamefont
  {J.}~\bibnamefont {Schilling}},\ }\href {\doibase 10.1002/adom.201400370}
  {\bibfield  {journal} {\bibinfo  {journal} {Advanced Optical Materials}\
  }\textbf {\bibinfo {volume} {3}},\ \bibinfo {pages} {129} (\bibinfo {year}
  {2015})}\BibitemShut {NoStop}%
\bibitem [{\citenamefont {Manganelli}\ \emph {et~al.}(2015)\citenamefont
  {Manganelli}, \citenamefont {Pintus},\ and\ \citenamefont
  {Bonati}}]{Manganelli2015}%
  \BibitemOpen
  \bibfield  {author} {\bibinfo {author} {\bibfnamefont {C.~L.}\ \bibnamefont
  {Manganelli}}, \bibinfo {author} {\bibfnamefont {P.}~\bibnamefont {Pintus}},
  \ and\ \bibinfo {author} {\bibfnamefont {C.}~\bibnamefont {Bonati}},\ }\href
  {\doibase 10.1364/OE.23.028649} {\bibfield  {journal} {\bibinfo  {journal}
  {Opt. Express}\ }\textbf {\bibinfo {volume} {23}},\ \bibinfo {pages} {28649}
  (\bibinfo {year} {2015})}\BibitemShut {NoStop}%
\bibitem [{\citenamefont {Azadeh}\ \emph {et~al.}(2015)\citenamefont {Azadeh},
  \citenamefont {Merget}, \citenamefont {Nezhad},\ and\ \citenamefont
  {Witzens}}]{Azadeh2015}%
  \BibitemOpen
  \bibfield  {author} {\bibinfo {author} {\bibfnamefont {S.~S.}\ \bibnamefont
  {Azadeh}}, \bibinfo {author} {\bibfnamefont {F.}~\bibnamefont {Merget}},
  \bibinfo {author} {\bibfnamefont {M.~P.}\ \bibnamefont {Nezhad}}, \ and\
  \bibinfo {author} {\bibfnamefont {J.}~\bibnamefont {Witzens}},\ }\href
  {https://www.osapublishing.org/ol/abstract.cfm?uri=ol-40-8-1877} {\bibfield
  {journal} {\bibinfo  {journal} {Optics letters}\ }\textbf {\bibinfo {volume}
  {40}},\ \bibinfo {pages} {1877} (\bibinfo {year} {2015})}\BibitemShut
  {NoStop}%
\bibitem [{\citenamefont {Sharma}\ \emph {et~al.}(2015)\citenamefont {Sharma},
  \citenamefont {Puckett}, \citenamefont {Lin}, \citenamefont {Vallini},\ and\
  \citenamefont {Fainman}}]{Sharma2015}%
  \BibitemOpen
  \bibfield  {author} {\bibinfo {author} {\bibfnamefont {R.}~\bibnamefont
  {Sharma}}, \bibinfo {author} {\bibfnamefont {M.~W.}\ \bibnamefont {Puckett}},
  \bibinfo {author} {\bibfnamefont {H.-H.}\ \bibnamefont {Lin}}, \bibinfo
  {author} {\bibfnamefont {F.}~\bibnamefont {Vallini}}, \ and\ \bibinfo
  {author} {\bibfnamefont {Y.}~\bibnamefont {Fainman}},\ }\href {\doibase
  10.1063/1.4922734} {\bibfield  {journal} {\bibinfo  {journal} {Applied
  Physics Letters}\ }\textbf {\bibinfo {volume} {106}},\ \bibinfo {pages}
  {241104} (\bibinfo {year} {2015})}\BibitemShut {NoStop}%
\bibitem [{\citenamefont {Harrison}(1989)}]{HarrisonBook1989}%
  \BibitemOpen
  \bibfield  {author} {\bibinfo {author} {\bibfnamefont {W.~A.}\ \bibnamefont
  {Harrison}},\ }\href@noop {} {\emph {\bibinfo {title} {Electronic Structure
  and the Properties of Solids}}},\ Dover Books on Physics\ (\bibinfo
  {publisher} {Dover Publications},\ \bibinfo {year} {1989})\BibitemShut
  {NoStop}%
\bibitem [{\citenamefont {Harrison}(1973{\natexlab{a}})}]{Harrison1973}%
  \BibitemOpen
  \bibfield  {author} {\bibinfo {author} {\bibfnamefont {W.}~\bibnamefont
  {Harrison}},\ }\href {http://prb.aps.org/abstract/PRB/v8/i10/p4487\_1}
  {\bibfield  {journal} {\bibinfo  {journal} {Physical Review B}\ }\textbf
  {\bibinfo {volume} {8}} (\bibinfo {year} {1973}{\natexlab{a}})}\BibitemShut
  {NoStop}%
\bibitem [{\citenamefont {Harrison}\ and\ \citenamefont
  {Phillips}(1974)}]{Harrison1974b}%
  \BibitemOpen
  \bibfield  {author} {\bibinfo {author} {\bibfnamefont {W.}~\bibnamefont
  {Harrison}}\ and\ \bibinfo {author} {\bibfnamefont {J.}~\bibnamefont
  {Phillips}},\ }\href {http://link.aps.org/doi/10.1103/PhysRevLett.33.410
  http://journals.aps.org/prl/abstract/10.1103/PhysRevLett.33.410} {\bibfield
  {journal} {\bibinfo  {journal} {Physical Review Letters}\ }\textbf {\bibinfo
  {volume} {33}},\ \bibinfo {pages} {6} (\bibinfo {year} {1974})}\BibitemShut
  {NoStop}%
\bibitem [{\citenamefont {Harrison}(1973{\natexlab{b}})}]{Harrison1973a}%
  \BibitemOpen
  \bibfield  {author} {\bibinfo {author} {\bibfnamefont {W.}~\bibnamefont
  {Harrison}},\ }\href {http://pra.aps.org/abstract/PRA/v7/i6/p1876\_1}
  {\bibfield  {journal} {\bibinfo  {journal} {Physical Review A}\ }\textbf
  {\bibinfo {volume} {7}},\ \bibinfo {pages} {1876} (\bibinfo {year}
  {1973}{\natexlab{b}})}\BibitemShut {NoStop}%
\bibitem [{\citenamefont {Huang}(1975)}]{Huang1975}%
  \BibitemOpen
  \bibfield  {author} {\bibinfo {author} {\bibfnamefont {C.}~\bibnamefont
  {Huang}},\ }\href@noop {} {\  (\bibinfo {year} {1975})}\BibitemShut {NoStop}%
\bibitem [{\citenamefont {Landau}\ and\ \citenamefont
  {Lifshitz}(ress)}]{Landau7}%
  \BibitemOpen
  \bibfield  {author} {\bibinfo {author} {\bibfnamefont {L.~D.}\ \bibnamefont
  {Landau}}\ and\ \bibinfo {author} {\bibfnamefont {E.~M.}\ \bibnamefont
  {Lifshitz}},\ }\href@noop {} {\emph {\bibinfo {title} {The Theory of
  Elasticity}}},\ Vol.~\bibinfo {volume} {7}\ (\bibinfo  {publisher}
  {required},\ \bibinfo {year} {Pergamon Press})\BibitemShut {NoStop}%
\bibitem [{\citenamefont {Bir}\ and\ \citenamefont {Pieu}(1972)}]{BirSymmetry}%
  \BibitemOpen
  \bibfield  {author} {\bibinfo {author} {\bibfnamefont {G.~L.}\ \bibnamefont
  {Bir}}\ and\ \bibinfo {author} {\bibfnamefont {G.~E.}\ \bibnamefont {Pieu}},\
  }\href@noop {} {\emph {\bibinfo {title} {Symmetry and Strain-Induced Effects
  in Semiconductors}}}\ (\bibinfo  {publisher} {John Willey and Sons},\
  \bibinfo {address} {New York/Toronto},\ \bibinfo {year} {1972})\BibitemShut
  {NoStop}%
\bibitem [{\citenamefont {Jha}\ and\ \citenamefont
  {Bloembergen}(1968)}]{Jha1968}%
  \BibitemOpen
  \bibfield  {author} {\bibinfo {author} {\bibfnamefont {S.}~\bibnamefont
  {Jha}}\ and\ \bibinfo {author} {\bibfnamefont {N.}~\bibnamefont
  {Bloembergen}},\ }\href {http://prola.aps.org/abstract/PR/v171/i3/p891\_1}
  {\bibfield  {journal} {\bibinfo  {journal} {Physical Review}\ }\textbf
  {\bibinfo {volume} {171}} (\bibinfo {year} {1968})}\BibitemShut {NoStop}%
\bibitem [{\citenamefont {Phillips}(1968)}]{Philips1968a}%
  \BibitemOpen
  \bibfield  {author} {\bibinfo {author} {\bibfnamefont {J.}~\bibnamefont
  {Phillips}},\ }\href {http://prola.aps.org/abstract/PR/v168/i3/p905\_1}
  {\bibfield  {journal} {\bibinfo  {journal} {Physical Review}\ }\textbf
  {\bibinfo {volume} {168}},\ \bibinfo {pages} {905} (\bibinfo {year}
  {1968})}\BibitemShut {NoStop}%
\bibitem [{\citenamefont {Damas}\ \emph {et~al.}(2013)\citenamefont {Damas},
  \citenamefont {Roux}, \citenamefont {Cassan}, \citenamefont {Marris-morini},
  \citenamefont {Izard}, \citenamefont {Bosseboeuf}, \citenamefont {Maroutian},
  \citenamefont {Lecoeur},\ and\ \citenamefont {Vivien}}]{Damas2013}%
  \BibitemOpen
  \bibfield  {author} {\bibinfo {author} {\bibfnamefont {P.}~\bibnamefont
  {Damas}}, \bibinfo {author} {\bibfnamefont {X.~L.}\ \bibnamefont {Roux}},
  \bibinfo {author} {\bibfnamefont {E.}~\bibnamefont {Cassan}}, \bibinfo
  {author} {\bibfnamefont {D.}~\bibnamefont {Marris-morini}}, \bibinfo {author}
  {\bibfnamefont {N.}~\bibnamefont {Izard}}, \bibinfo {author} {\bibfnamefont
  {A.}~\bibnamefont {Bosseboeuf}}, \bibinfo {author} {\bibfnamefont
  {T.}~\bibnamefont {Maroutian}}, \bibinfo {author} {\bibfnamefont
  {P.}~\bibnamefont {Lecoeur}}, \ and\ \bibinfo {author} {\bibfnamefont
  {L.}~\bibnamefont {Vivien}}\ }(\bibinfo {year} {2013})\ pp.\ \bibinfo {pages}
  {11--13}\BibitemShut {NoStop}%
\bibitem [{\citenamefont {Schriever}\ \emph {et~al.}(2010)\citenamefont
  {Schriever}, \citenamefont {Bohley},\ and\ \citenamefont
  {Wehrspohn}}]{Schriever2010}%
  \BibitemOpen
  \bibfield  {author} {\bibinfo {author} {\bibfnamefont {C.}~\bibnamefont
  {Schriever}}, \bibinfo {author} {\bibfnamefont {C.}~\bibnamefont {Bohley}}, \
  and\ \bibinfo {author} {\bibfnamefont {R.~B.}\ \bibnamefont {Wehrspohn}},\
  }\href {http://www.ncbi.nlm.nih.gov/pubmed/20125692} {\bibfield  {journal}
  {\bibinfo  {journal} {Optics letters}\ }\textbf {\bibinfo {volume} {35}},\
  \bibinfo {pages} {273} (\bibinfo {year} {2010})}\BibitemShut {NoStop}%
\bibitem [{\citenamefont {Witzens}(2014)}]{Witzens2014}%
  \BibitemOpen
  \bibfield  {author} {\bibinfo {author} {\bibfnamefont {J.}~\bibnamefont
  {Witzens}},\ }\href {\doibase 10.1016/j.cpc.2014.03.016} {\bibfield
  {journal} {\bibinfo  {journal} {Computer Physics Communications}\ }\textbf
  {\bibinfo {volume} {185}},\ \bibinfo {pages} {2221} (\bibinfo {year}
  {2014})}\BibitemShut {NoStop}%
\bibitem [{\citenamefont {Appelbaum}\ and\ \citenamefont
  {Hamann}(1973)}]{Appelbaum1973}%
  \BibitemOpen
  \bibfield  {author} {\bibinfo {author} {\bibfnamefont {J.~a.}\ \bibnamefont
  {Appelbaum}}\ and\ \bibinfo {author} {\bibfnamefont {D.~R.}\ \bibnamefont
  {Hamann}},\ }\href@noop {} {\bibfield  {journal} {\bibinfo  {journal}
  {Physical Review B}\ }\textbf {\bibinfo {volume} {8}},\ \bibinfo {pages}
  {1777} (\bibinfo {year} {1973})}\BibitemShut {NoStop}%
\bibitem [{\citenamefont {Anderson}(1969)}]{Anderson1969}%
  \BibitemOpen
  \bibfield  {author} {\bibinfo {author} {\bibfnamefont {P.~W.}\ \bibnamefont
  {Anderson}},\ }\href {\doibase 10.1103/PhysRev.181.25} {\bibfield  {journal}
  {\bibinfo  {journal} {Physical Review}\ }\textbf {\bibinfo {volume} {181}},\
  \bibinfo {pages} {25} (\bibinfo {year} {1969})}\BibitemShut {NoStop}%
\bibitem [{\citenamefont {Chelikowsky}\ \emph {et~al.}(1991)\citenamefont
  {Chelikowsky}, \citenamefont {Glassford},\ and\ \citenamefont
  {Phillips}}]{Chelikowsky1991}%
  \BibitemOpen
  \bibfield  {author} {\bibinfo {author} {\bibfnamefont {J.~R.}\ \bibnamefont
  {Chelikowsky}}, \bibinfo {author} {\bibfnamefont {K.~M.}\ \bibnamefont
  {Glassford}}, \ and\ \bibinfo {author} {\bibfnamefont {J.~C.}\ \bibnamefont
  {Phillips}},\ }\href {http://prb.aps.org/abstract/PRB/v44/i4/p1538\_1}
  {\bibfield  {journal} {\bibinfo  {journal} {Physical Review B}\ }\textbf
  {\bibinfo {volume} {44}},\ \bibinfo {pages} {1538} (\bibinfo {year}
  {1991})}\BibitemShut {NoStop}%
\bibitem [{\citenamefont {Wendel}\ and\ \citenamefont
  {Martin}(1978)}]{Wendel1978}%
  \BibitemOpen
  \bibfield  {author} {\bibinfo {author} {\bibfnamefont {H.}~\bibnamefont
  {Wendel}}\ and\ \bibinfo {author} {\bibfnamefont {R.~M.}\ \bibnamefont
  {Martin}},\ }\href
  {http://journals.aps.org/prl/abstract/10.1103/PhysRevLett.40.950} {\bibfield
  {journal} {\bibinfo  {journal} {Physical Review Letters}\ }\textbf {\bibinfo
  {volume} {40}},\ \bibinfo {pages} {950} (\bibinfo {year} {1978})}\BibitemShut
  {NoStop}%
\bibitem [{\citenamefont {Phillips}\ and\ \citenamefont
  {Kleinman}(1959)}]{Phillips1959}%
  \BibitemOpen
  \bibfield  {author} {\bibinfo {author} {\bibfnamefont {J.~C.}\ \bibnamefont
  {Phillips}}\ and\ \bibinfo {author} {\bibfnamefont {L.}~\bibnamefont
  {Kleinman}},\ }\href {\doibase 10.1103/PhysRev.116.287} {\bibfield  {journal}
  {\bibinfo  {journal} {Physical Review}\ }\textbf {\bibinfo {volume} {116}},\
  \bibinfo {pages} {287} (\bibinfo {year} {1959})}\BibitemShut {NoStop}%
\bibitem [{\citenamefont {Borghi}\ \emph {et~al.}(2015)\citenamefont {Borghi},
  \citenamefont {Mancinelli}, \citenamefont {Merget}, \citenamefont {Witzens},
  \citenamefont {Bernard}, \citenamefont {Ghulinyan}, \citenamefont {Pucker},\
  and\ \citenamefont {Pavesi}}]{Borghi2015b}%
  \BibitemOpen
  \bibfield  {author} {\bibinfo {author} {\bibfnamefont {M.}~\bibnamefont
  {Borghi}}, \bibinfo {author} {\bibfnamefont {M.}~\bibnamefont {Mancinelli}},
  \bibinfo {author} {\bibfnamefont {F.}~\bibnamefont {Merget}}, \bibinfo
  {author} {\bibfnamefont {J.}~\bibnamefont {Witzens}}, \bibinfo {author}
  {\bibfnamefont {M.}~\bibnamefont {Bernard}}, \bibinfo {author} {\bibfnamefont
  {M.}~\bibnamefont {Ghulinyan}}, \bibinfo {author} {\bibfnamefont
  {G.}~\bibnamefont {Pucker}}, \ and\ \bibinfo {author} {\bibfnamefont
  {L.}~\bibnamefont {Pavesi}},\ }\href {\doibase 10.1364/OL.40.005287}
  {\bibfield  {journal} {\bibinfo  {journal} {Opt. Lett.}\ }\textbf {\bibinfo
  {volume} {40}},\ \bibinfo {pages} {5287} (\bibinfo {year}
  {2015})}\BibitemShut {NoStop}%
\end{thebibliography}%
\end{document}